\documentclass[twocolumn, superscriptaddress,prd,amsmath,amsfonts,nofootinbib,showpacs]{revtex4-1}

\newcommand{\ud}{\text{d}}

\newcommand{\half}{{\tfrac{1}{2}}}
\newcommand{\gb}{\check{g}}
\newcommand{\bs}{\begin{split}}
\newcommand{\es}{\end{split}}
\newcommand{\delb}{\check{\nabla}}

\newcommand{\mc}[1]{\mathcal{#1}}
\newcommand{\bc}[1]{\check{\mathcal{#1}}}
\newcommand{\TT}{\textsc{tt}}
\newcommand{\hb}{\bar{h}}

\usepackage{float}
\usepackage{color}
\usepackage{url}
\usepackage[ocgcolorlinks]{hyperref}
\hypersetup{
    colorlinks,
    citecolor=blue,
    filecolor=blue,
    linkcolor=blue,
    urlcolor=blue,
    pdftitle={Localizing the Angular Momentum of Linear Gravity},
    pdfauthor={Luke M. Butcher}}
\usepackage{slashed}
\usepackage{graphicx}
\usepackage{subfigure}
\usepackage{psfrag}
\usepackage{amsthm}
\theoremstyle{plain}

\begin{document}

\title{Localizing the Angular Momentum of Linear Gravity}
\author{Luke M. Butcher}
\email[]{l.butcher@mrao.cam.ac.uk}
\affiliation{Astrophysics Group, Cavendish Laboratory, J J Thomson Avenue, Cambridge, CB3 0HE, UK}
\affiliation{Kavli Institute for Cosmology, Madingley Road, Cambridge, CB3 0HA, UK}
\author{Anthony Lasenby}
\affiliation{Astrophysics Group, Cavendish Laboratory, J J Thomson Avenue, Cambridge, CB3 0HE, UK}
\affiliation{Kavli Institute for Cosmology, Madingley Road, Cambridge, CB3 0HA, UK}
\author{Michael Hobson}
\affiliation{Astrophysics Group, Cavendish Laboratory, J J Thomson Avenue, Cambridge, CB3 0HE, UK}
\date{\today}

\begin{abstract}
In a previous article \cite{Butcher10} we derived an energy-momentum tensor for linear gravity that exhibited positive energy-density and causal energy-flux. Here we extend this framework by localizing the angular momentum of the linearized gravitational field, deriving a gravitational spin tensor which possesses similarly desirable properties.  By examining the local exchange of angular momentum (between matter and gravity) we find that gravitational intrinsic spin is localized, separately from ``orbital'' angular momentum, in terms of a gravitational spin tensor. This spin tensor is then uniquely determined by requiring that it obey two simple physically-motivated algebraic conditions.  Firstly, the spin of an arbitrary (harmonic-gauge) gravitational plane-wave is required to flow in the direction of propagation of the wave. Secondly, the spin tensor of any transverse-traceless gravitational field is required to be traceless. (The second condition ensures that \emph{local} field redefinitions suffice to cast our gravitational energy-momentum tensor and spin tensor as sources of gravity in a quadratic approximation to general relativity, and simultaneously rids the gravitational field of infinite pressure gradients). Additionally, the following properties arise in the spin tensor spontaneously: all transverse-traceless fields have purely spatial spin, and any field generated by a static distribution of matter will carry no spin at all. Following the structure of our previous paper, we then examine the (spatial) angular momentum exchanged between the gravitational field and an infinitesimal detector, and develop a microaveraging procedure that renders the process gauge-invariant. The exchange of non-spatial angular momentum (i.e.\ moment-of-energy) is also analysed, leading us to conclude that a gravitational wave can displace the centre-of-mass of the detector; this conclusion is also confirmed by a ``first principles'' treatment of the system. Finally, we discuss the spin carried by a gravitational plane-wave.

\end{abstract}
\maketitle

\section{Introduction}\label{intro}
We recently developed a local description of energy and momentum in linear gravity, deriving a gravitational energy-momentum tensor $\tau_{ab}$ that describes positive energy-density and causal energy-flux \cite{Butcher10}.\footnote{To clarify: we have not performed the demonstrably \emph{impossible} feat of finding a tensor $\tau_{ab}$, quadratic in the gravitational field $h_{ab}$ and second-order in derivatives, that is invariant under the linearized gauge transformation $\delta h_{ab}=\delb_{(a}\xi_{b)}$. Rather, we rely on a gauge-fixing programme (motivated by key properties of $\tau_{ab}$ and the energetics of an infinitesimal gravitational detector) to remove the freedom to perform such transformations, and hence arrive at a physically unambiguous description. } The purpose of this present article is to complete our picture of local linear gravitational energetics, extending our framework to quantify the \emph{angular momentum} carried by the field. This approach will localize both the ``orbital'' angular momentum and the intrinsic spin of linear gravity, the former in terms of $\tau_{ab}$, and the latter in terms of a gravitational spin tensor $s^{a}_{\phantom{a}bc}$. Not only is this spin tensor vital if one is to account for the angular momentum possessed by gravity and exchanged locally with matter, the formula we derive for it will display a number physically desirable algebraic properties, closely analogous to those of $\tau_{ab}$.

Armed with a local description of the energy, momentum, and angular momentum of linear gravity, we will be ready to tackle the task of our next paper \cite{Ang2}: to understand $\tau_{ab}$ and $s^{a}_{\phantom{a}bc}$ in terms of the familiar theoretical apparatus that has been used to define gravitational energy-momentum in the past \cite{Einstein,LL, ADMII}, and energy-momentum in general \cite{noether,hilbert,belinfante}. These developments will crystallize the tensors' physical interpretation, deepen our understanding of their theoretical underpinnings, and suggest a route by which our work might be generalized beyond the linear approximation.

Let us begin by summarizing the key points of the program developed in \cite{Butcher10}.\footnote{As before, we work in units where $c=1$, write $\kappa \equiv 8\pi G$, and use the sign conventions of Wald \cite{Wald}: $\eta_{\mu\nu}\equiv \mathrm{diag}(-1,1,1,1)$, $[\nabla_c,\nabla_d]v^a\equiv 2 \nabla_{[c}\nabla_{d]}v^a  \equiv R^{a}_{\phantom{a}bcd}v^b$, and $R_{ab}\equiv R^{c}_{\phantom{c}acb}$. We use Roman letters (except $i,j,k,l$) as abstract tensor indices \citep[chap.\ 2.4]{Wald} and Greek letters as numerical indices running from 0 to 3. The indices $i,j,k,l$ are reserved for spatial components, and run from 1 to 3.} We define the gravitational field $h_{ab}$ on a \emph{flat} background spacetime $(\bc{M},\gb_{ab})$ by a diffeomorphism $\phi: \mc{M} \to \bc{M}$ that maps the physical spacetime $(\mc{M},g_{ab})$ onto the background:\footnote{As usual, fields defined on $\mc{M}$ have their indices raised and lowered with $g_{ab}$, and those on $\bc{M}$ with $\gb_{ab}$. Lorentzian coordinates $\{x^\mu\}$ are commonly deployed in $\bc{M}$, for which $\gb_{\mu\nu} =\eta_{\mu\nu}$.}
\begin{align}\label{hdef}
\phi^* g_{ab}= \gb_{ab} + h_{ab}.
\end{align}
The physical spacetime is assumed to be ``nearly flat'', and $\phi$ chosen such that  $h_{ab}$ is small everywhere, so that the linear approximation to the Einstein field equations is valid:
\begin{align}\label{FEqs}
\widehat{G}_{ab}^{\phantom{ab}cd}h_{cd}= \kappa \check{T}_{ab} + O(h^2),
\end{align}
where $\check{T}_{ab}\equiv \phi^* T_{ab}\sim O(h)$ is the matter energy-momentum tensor $T_{ab}$ mapped onto the background, and
\begin{align}\nonumber
\widehat{G}_{ab}^{\phantom{ab}cd}h_{cd} &\equiv \delb_c \delb_{(a}h_{b)}^{\phantom{a)}c} - \half \delb^2 h_{ab} - \half\delb_a\delb_b h
\\\label{Gdef}&\quad+ \half\gb_{ab}\left(\delb^2 h - \delb_c\delb_d h^{cd} \right)
\end{align}
is the linearized Einstein tensor $G^{(1)}_{ab}$.

The gravitational energy-momentum tensor $\tau_{ab}$ is defined by seeking a symmetric tensor, quadratic in $\delb_{c}h_{ab}$, which solves
\begin{align}\label{exchange}
\delb_a j_\mu^{\phantom{\mu}a} + \phi^*(\nabla_a J_\mu^{\phantom{\mu}a}) =0,
\end{align}
neglecting terms $O(h^3)$.  In the above equation, $J_\mu^{\phantom{\mu}a} \equiv T^a_{\phantom{a}b}e_\mu^{\phantom{\mu}b}$ are the (1 energy, 3 momentum) current-densities of matter, associated with the (1 timelike, 3 spacelike) vector fields $e_\mu^{\phantom{\mu}a} \equiv (\phi^{-1})^*\check{e}_\mu^{\phantom{\mu}a}$, the images of the Lorentzian coordinate basis $\check{e}_\mu^{\phantom{\mu}a}\equiv (\partial/\partial x^\mu)^a$ that generate the translational symmetries of the background; the $j_\mu^{\phantom{\mu}a} \equiv \tau^a_{\phantom{a}b}\check{e}_\mu^{\phantom{\mu}b}= \tau^a_{\phantom{a}\mu}$ constitute the energy-momentum current-densities of the gravitational field. Consequently (\ref{exchange}) indicates that the extent to which material energy-momentum fails to be conserved at a point in the physical spacetime is exactly equal and opposite to the extent to which gravitational energy-momentum fails to be conserved at the corresponding point in the background. Interactions between matter and gravity can then be understood in terms of a \emph{local exchange} of energy and momentum between the two.

It is not possible to construct a $\tau_{ab}$ to solve (\ref{exchange}) for all gravitational fields, so a condition must be placed on $h_{ab}$ in order to proceed. Of all possible symmetric tensors $\tau_{ab}$, quadratic in $\delb_{c}h_{ab}$, and all (non-trivial, linear and Lorentz invariant) field conditions, \emph{only one} combination solves (\ref{exchange}):
\begin{align}
\label{taubar}
\kappa \bar{\tau}_{ab}=\tfrac{1}{4}\delb_a  h_{cd}\delb_b \hb^{cd},
\end{align}
\begin{align}\label{harmonic}
\delb^a \hb_{ab} =0,
\end{align}
where the overbars signify trace-reversal. Because (\ref{harmonic}) is simply  the equation of \emph{harmonic gauge}, which can always be satisfied through a choice of $\phi$, the field condition does not restrict the physical applicability of our approach in any respect. In fact, the only effect of the field condition is to vastly reduce the gauge freedom in our description of gravitational energy-momentum (\ref{taubar}). What at first appeared as a weakness, is in fact a great strength of our approach. Essentially, (\ref{harmonic}) indicates that $\phi$ is to be chosen such that it maps Lorentzian coordinates $\{x^\mu\}$ of the background onto harmonic coordinates $y^\mu(p) \equiv x^\mu(\phi(p))$ of the physical spacetime. This ensures that the energy-momentum currents $J_\mu^{\phantom{\mu}a} $ are defined by the generators of a harmonic coordinate system; these represent the approximate translational symmetries of the physical spacetime (present due to its small curvature) and give a sensible replacement for killing vectors in the absence of an exact symmetry.

The gravitational energy-momentum tensor $\tau_{ab}$ has two notable mathematical properties, in addition to solving (\ref{exchange}). Firstly, the energy-momentum tensor for any (harmonic gauge) gravitational plane-wave
\begin{align}\label{planewave}
h_{ab}=h_{ab}(x^\alpha k_\alpha),\qquad k^a \hb_{ab}=0,\qquad k^ak_a=0,
\end{align}
is completely invariant under the remaining gauge freedom consistent with (\ref{harmonic}) and (\ref{planewave}). Secondly, and most remarkably of all, $\tau_{ab}$ displays the following \emph{positivity property}: all transverse-traceless (\TT) gravitational fields have positive energy-density and causal energy-flux, for all observers. To state this rigorously: if, at some point $p\in \bc{M}$, the gravitational field $h_{ab}$ obeys the transverse-traceless conditions
\begin{align}\label{TTcon}
\delb^a h_{ab}=0, \quad h=0, \quad u^a h_{ab}=0,
\end{align}
for some timelike vector $u^a$, then $\tau_{ab}$ satisfies the following inequalities
\begin{align}\label{pos1}
v^a \tau_{ab} v^b &\ge 0,\\
\label{pos2}
v^a \tau_{ac}\tau^{c}_{\phantom{c}b} v^b &\le 0,
\end{align}
at $p$, for any timelike vector $v^a$.

In order to deal with the last trace of gauge freedom that remains after enforcing (\ref{harmonic}), we examined the energy-momentum transferred between the gravitational field and an infinitesimal probe, i.e.\ a matter ``point-source'' with energy-momentum tensor 
\begin{align}\nonumber
\check{T}_{00}&= M \delta(\vec{x}) + \half I_{ij} \partial_i \partial_j \delta(\vec{x}),\\\label{pointsource}
\check{T}_{0i}&= \half(\dot{I}_{ij} -L_{ij})\partial_j \delta (\vec{x}),\\\nonumber
\check{T}_{ij}&= \half \ddot{I}_{ij} \delta (\vec{x}),
\end{align}
derived by shrinking a compact source down to a point.\footnote{$M$, $I_{ij}$ and $L_{ij}$ are the mass, moment of inertia, and angular momentum of the source, respectively. Overdots indicate differentiation with respect to $t\equiv x^0$, and the three spatial coordinates are abbreviated $\vec{x}=(x^1,x^2,x^3)$. In \cite{Butcher10}, the angular momentum of matter was written $-J_{ij}$; our change in notation corrects for the unusual sign convention chosen in equation (A10) of \cite{Butcher10}, and avoids confusion with the energy-momentum current-densities $J_\mu^{\phantom{\mu}a}$.} The exchange is rendered gauge-invariant (under the freedom that remains after (\ref{harmonic}) has been enforced) by the \emph{monopole-free microaverage}: the  incoming wave is split into a sum of Heaviside step-functions, and the energy-momentum delivered by each is integrated over a vanishingly small 4-volume centered on the probe.\footnote{Details are to be found in Sections IV B, and IV C of \cite{Butcher10}.} The result
\begin{align}\label{MA=TT}
\left\langle\partial^\mu \tau_{\mu\nu}\right\rangle^\slashed{M}_{\int} = -\tfrac{1}{4} \delta(\vec{x})\ddot{I}_{ij}\partial_\nu h^\TT_{ij}
\end{align}
is equal to the bare (i.e.\ not microaveraged) energy-momentum delivered by the incident field in \TT-gauge. This motivated the program of fixing the final piece of gauge freedom by insisting that the incident $h_{ab}$ be transverse-traceless; consequently, $\tau_{ab}$ represents the gauge-invariant gravitational energy-momentum that is accessible to an infinitesimal probe at rest in the \TT\ frame. Furthermore, due to the positivity property of $\tau_{ab}$, this program ensures that the gravitational field is always described with positive energy-density and causal energy flux.

The approach we will take for localizing gravitational angular momentum will be very similar to the one we have just described. Section \ref{GAM} of this paper begins with the counterpart of (\ref{exchange}) for the local exchange of angular momentum  between matter and gravity. We will show that the local change in the angular momentum of matter is not entirely accounted for by the change in orbital angular momentum $2x_{[\mu}\tau_{\nu]}{}^a$ carried by the gravitational field: gravity's intrinsic spin $s^a_{\phantom{a}\mu\nu}$ must be included to balance the exchange. This argument defines $s^a_{\phantom{a}bc}$ up to the addition of total divergences, so further requirements must be placed on the tensor before we have a unique formula localizing gravitational intrinsic spin. We achieve this in section \ref{Spintensor} by demanding that $s^a_{\phantom{a}bc}$ satisfy two simple, physically motivated, algebraic conditions, analogous to the algebraic properties of $\tau_{ab}$. As a result, a formula (\ref{spintensor}) is derived for the spin tensor of the gravitational field. The gauge freedom of $s^a_{\phantom{a}bc}$ is automatically nullified by the \TT\ program motivated in \cite{Butcher10}; however, it is still enlightening to reprise our analysis of the infinitesimal probe and develop a microaverage procedure that renders the transfer of angular momentum gauge-invariant (within harmonic gauge) without the need to fix the gauge completely. This is covered in section \ref{Inter}. In section \ref{MoE} we examine the role of the non-spatial components of gravitational angular momentum, demonstrating that the exchange of non-spatial spin $s^a_{\phantom{a}0i}$ can displace the centre-of-mass of a gravitational probe. We conclude our investigation with a calculation and analysis of the intrinsic spin carried by a gravitational plane-wave.

\section{Local Angular Momentum Exchange}\label{GAM}
The purpose of this section, and the one that follows it, is to extend the basic framework of \cite{Butcher10} to include a local description of gravitational angular momentum. Unlike our work on $\tau_{ab}$, eliminating gauge freedom will not be a major concern: we already know that harmonic gauge (\ref{harmonic}) is necessary (thus we shall enforce this condition throughout the article), and that the last trace of freedom must be removed by insisting that the incident field be transverse-traceless. We begin by formulating the local exchange of angular momentum.

As noted in Section \ref{intro}, the material energy-momentum current-densities $J_\mu^{\phantom{\mu}a}$ are formed by contracting $T^a_{\phantom{a}b}$ with the vectors $e_\mu^{\phantom{\mu}b}\equiv (\phi^{-1})^*\check{e}_\mu^{\phantom{\mu}b}$, the pushforwards of which under $\phi$ generate the translational symmetries of the background. Therefore, to define material angular momentum current-densities $J_{\mu\nu}^{\phantom{\mu\nu}a}$, we must contract $T^a_{\phantom{a}b}$ with the vector fields $(\phi^{-1})^*(2x_{[\mu}\check{e}_{\nu]}{}^b)$, the pushforwards of which generate the \emph{rotational} symmetries of the background:\footnote{We use the term ``rotational symmetry'' here as a shorthand for both rotations and Lorentz boosts. The three independent vector fields $2x_{[i}\check{e}_{j]}{}^a$ generate rotations (so that $2x_{[1}\check{e}_{2]}{}^a$ rotates about the $x^3$-axis, for example) and hence define angular momentum current-densities. The $2x_{[0}\check{e}_{i]}{}^a$ generate boosts (in the $x^i$-direction) and define moment-of-energy current-densities, the interpretation of which we explore in section \ref{MoE}.}
\begin{align}\nonumber
J_{\mu\nu}^{\phantom{\mu\nu}a}&\equiv T^a_{\phantom{a}b} (\phi^{-1})^*(2x_{[\mu}\check{e}_{\nu]}{}^b)\\\label{matterJ}
&=  2T^a_{\phantom{a}b} y_{[\mu}e_{\nu]}{}^b. 
\end{align}
As usual, $\{x^\mu\}$ comprise a Lorentzian coordinate system on the background, and $y^\mu(p) \equiv x^\mu(\phi(p))$ are the image of these coordinates in the physical spacetime. The  $\{y^\mu\}$ are harmonic (that is, $\nabla^2 y^\mu=0$) as a result of the gauge condition (\ref{harmonic}). 

We wish to explain the effect of the gravitational field on the angular momentum of matter in terms of a local exchange of angular momentum between the two. Just as (\ref{exchange}) captured this idea for energy-momentum, we will require
\begin{align}\label{AMexchange}
\delb_a j_{\mu\nu}^{\phantom{\mu\nu}a} + \phi^*(\nabla_a J_{\mu\nu}^{\phantom{\mu\nu}a}) =0
\end{align}
for angular momentum, where $j_{\mu\nu}^{\phantom{\mu\nu}a}$ is the angular momentum current-density of the gravitational field. Neglecting terms $O(h^3)$, equation (\ref{AMexchange}) is equivalent to
\begin{align}\nonumber
\delb_a j_{\mu\nu}^{\phantom{\mu\nu}a} &= -\phi^*(\nabla_a(J^{\phantom{\mu\nu}\!\!a}_{\mu\nu}))\\\nonumber
&=  -\phi^*( T^a_{\phantom{a}b} \nabla_a (2y_{[\mu}e_{\nu]}{}^b))\\\nonumber
&= -\phi^*(T^a_{\phantom{a}b})\left[ (\delb_c h_a^{\phantom{a}b} +\delb_a h_c^{\phantom{c}b} - \delb^b h_{ac}  )x_{[\mu}\check{e}_{\nu]}{}^c \right.\\ \nonumber
&\qquad\!\qquad\qquad\left.{} +2\delb_a (x_{[\mu}\check{e}_{\nu]}{}^b) \right]\\\nonumber
&= -\check{T}^a_{\phantom{a}b}(\delb_c h_{a}^{\phantom{a}b})x_{[\mu} \check{e}_{\nu]}{}^c \\
& \quad-2(\check{T}^a_{\phantom{a}b}  - h^{ac}\check{T}_{cb}) \delb_{a}(x_{[\mu} \check{e}_{\nu]}{}^b),
\end{align}
where in the last line we used $\phi^*( T^a_{\phantom{a}b}) = \phi^*( g^{ac}T_{cb})= \check{T}^a_{\phantom{a}b} - h^{ac}\check{T}_{cb} + O(h^3)$ and $\check{T}_{ab}=\check{T}_{ba}$. As we now have an equation relating tensors defined on the background, we can express these tensors in terms of their components in the Lorentzian coordinate system:
\begin{align}\nonumber
\partial_\alpha j_{\mu\nu}^{\phantom{\mu\nu}\alpha} &= -\check{T}^\alpha_{\phantom{\alpha}\beta}(\partial_\gamma h_{\alpha}^{\phantom{\alpha}\beta})x_{[\mu} \delta_{\nu]}^\gamma \\\nonumber
&\ \!\quad-2(\check{T}^\alpha_{\phantom{\alpha}\beta}  - h^{\alpha\gamma}\check{T}_{\gamma\beta}) \partial_{\alpha}(x_{[\mu} \delta_{\nu]}^\beta)
\\\label{label1}
&= - x_{[\mu}\check{T}^{\alpha\beta} \partial_{\nu]}h_{\alpha\beta} + 2 h_{\beta[\mu}\check{T}_{\nu]}{}^\beta.
\end{align}
Finally, we recall the field equations (\ref{FEqs}) in harmonic gauge,
\begin{align}
  \partial^2 \hb_{ab}=-2 \kappa\check{T}_{\alpha\beta},
\end{align}
and eliminate $\check{T}_{\alpha\beta}$ from (\ref{label1}):
\begin{align}\nonumber
\partial_\alpha j_{\mu\nu}^{\phantom{\mu\nu}\alpha} &=  (\partial^2 \hb^{\alpha\beta}x_{[\mu}\partial_{\nu]}h_{\alpha\beta} -  2h_{\beta[\mu}\partial^2 \hb_{\nu]}{}^\beta)/2\kappa\\\label{sappears}
&=  \partial_\alpha\left[2x_{[\mu}\tau_{\nu]}{}^\alpha + h_{\beta[\nu}\partial^\alpha h_{\mu]}{}^\beta /\kappa\right].
\end{align}
This is rather surprising result, and one that reveals the importance of the gravitational field's \emph{intrinsic spin}. The first term in the square brackets clearly represents the \emph{orbital angular momentum} of the field: it takes the familiar form $x \times p$ and is the result of the tangential linear momentum about the origin. The second term, in contrast, does not depend explicitly on $x^\mu$; it measures the extent to which the field itself is spinning at a particular point, and contributes the same gravitational angular momentum without regard to \emph{where} this spin is taking place. We are forced by (\ref{sappears}) to accept that the angular momentum of the gravitational field is not simply orbital, but also has an \emph{intrinsic} component:
\begin{align}\label{t+s}
j_{\mu\nu}^{\phantom{\mu\nu}\alpha} = 2x_{[\mu}\tau_{\nu]}{}^\alpha + s^\alpha_{\phantom{\alpha}\mu\nu},
\end{align}
where $s^\alpha_{\phantom{\alpha}\mu\nu}$ is the gravitational \emph{spin tensor} (composed of intrinsic spin current-densities) without which the local exchange of angular momentum would not balance. Of course, the division of angular momentum into orbital and intrinsic components is not a new idea, and the form of equation (\ref{t+s}) originates from standard flat-space field theory \cite{Blag, Bogo}. In general, the Noether current of a rotational symmetry cannot be constructed entirely from Noether currents of translational symmetries: the mismatch, born of the  field's tensorial (or spinorial) structure, is called intrinsic spin.\footnote{Essentially, this is because a tensor field undergoes two types of transformation when it is rotated. A vector field $A^\mu (x)$, for example, becomes $\Lambda^\mu_{\phantom{\mu}\nu} A^\nu(\Lambda^{-1}(x))$; in the parlance of quantum field theory, this can be understood as a displacement $x\to \Lambda(x)$ generated by the orbital angular momentum operator $x \times p$, and a pointwise Lorentz transformation $A^\mu \to \Lambda^\mu_{\phantom{\mu}\nu} A^\nu$ generated by the spin operator.} More neatly, and of greater relevance to our later analysis, the energy-momentum tensor and the spin tensor can be derived separately from a Lagrangian by `gauging' the translational and rotational symmetries of spacetime and taking the functional derivatives with respect to the two gauge fields. In the paper that follows \cite{Ang2} we construct $\tau_{\mu\nu}$ and $s^\alpha_{\phantom{\alpha}\mu\nu}$ according to this method, confirming that our formulae for $\tau_{\mu\nu}$ and $s^\alpha_{\phantom{\alpha}\mu\nu}$ (soon to be derived) are in keeping with the established concepts of energy-momentum and spin.\footnote{Because spin tensors are usually associated with \emph{asymmetric} energy-momentum tensors, it is worth mentioning that the symmetry of $\tau_{\mu\nu}$ does not contradict the existence of $s^\alpha_{\phantom{\alpha}\mu\nu}$. Typically, one argues that $\tau_{[\mu\nu]}\ne 0$ describes finite torques acting on infinitessimal regions \cite{MTWcube}, and then states that this is only acceptable if one can interpret these torques as generating intrinsic spin: $\partial_\alpha s^\alpha_{\phantom{\alpha}\mu\nu}=2\tau_{[\mu\nu]}$. Clearly, this argument does not run in reverse: the presence of a spin tensor does not require that the energy-momentum tensor be asymmetric. A symmetric gravitational energy-momentum tensor simply indicates that there are no torques on infinitesimal regions due to gravity, and so (in the absence of matter) the spin-tensor is conserved: $\partial_\alpha s^\alpha_{\phantom{\alpha}\mu\nu}=2\tau_{[\mu\nu]}=0$.}

\section{Gravitational Intrinsic Spin Tensor}\label{Spintensor}
Our immediate goal, of course, is to arrive at a formula for $s^\alpha_{\phantom{\alpha}\mu\nu}$ in terms of $h_{\alpha\beta}$. With this in mind, it is tempting to solve (\ref{sappears}) simply by setting
\begin{align}\label{sguess}
\kappa s^\alpha_{\phantom{\alpha}\mu\nu} \stackrel{?}{\equiv}  h_{\beta[\nu}\partial^\alpha h_{\mu]}{}^\beta,
\end{align}
and declare that we have found our local description of gravitational spin. However, this is not the only solution: the exchange equation (\ref{sappears}) only defines $s^\alpha_{\phantom{\alpha}\mu\nu}$ up to terms with identically vanishing divergence, so further demands must be made of the spin tensor before it can be determined uniquely. Obviously, $s^\alpha_{\phantom{\alpha}\mu\nu}$ should have the same basic properties as $\tau_{\mu\nu}$: it should be a local, quadratic, Lorentz-covariant function of $h_{\alpha\beta}$, and contain no dimensionful constants other than $\kappa$.\footnote{This last stipulation (which forces the terms in $s^\alpha_{\phantom{\alpha}\mu\nu}$ to contain exactly one derivative, in order that they have the correct units) is essentially unavoidable within the context of classical general relativity: $\kappa$ is the only dimensionful constant available. If we allow ourselves to use Planck's constant $\hbar$ (as we would for a quantum theory) or introduce a new dimensionful gravitational constant (as would arise in a higher-derivative theory of gravity) then higher derivative terms would be dimensionally permissible within the spin tensor; nonetheless, these higher-derivative terms would each be multiplied by small factors (such as the Planck length) that would ensure the terms were negligible within the low-curvature regime of the theory that corresponds to classical general relativity.} The general solution to (\ref{sappears}) is then
\begin{align}\label{SP1}
\kappa s^\alpha_{\phantom{\alpha}\mu\nu} = h_{\beta[\nu}\partial^\alpha h_{\mu]}{}^\beta + \partial_\beta \Sigma^{\alpha\beta}{}_{\mu\nu},
\end{align}
where $\Sigma^{\alpha\beta}{}_{\mu\nu}$ is any local, quadratic, Lorentz-covariant function of $h_{\alpha\beta}$ (but not its derivatives) that obeys
\begin{align}
\Sigma^{\alpha\beta}{}_{\mu\nu}=-\Sigma^{\beta\alpha}{}_{\mu\nu}=-\Sigma^{\alpha\beta}{}_{\nu\mu}.
\end{align}
The most general tensor that can be formed from $h_{\alpha\beta}$  this way is 
\begin{align}\nonumber
\Sigma^{\alpha\beta}{}_{\mu\nu}&\equiv A_1 h^\alpha_{[\mu}h_{\nu]}^\beta + A_2 h h^{[\alpha}_{[\mu}\delta^{\beta]}_{\nu]}+A_3 h^\gamma_{[\mu}\delta_{\nu]}^{[\beta}h^{\alpha]}_\gamma \\\label{SP2}
&\quad {} + \delta^{\alpha}_{[\mu}\delta^{\beta}_{\nu]}\left(A_4 h^2 + A_5 h_{\gamma\delta}h^{\gamma\delta}\right),
\end{align}
where the $\{A_n\}$ are arbitrary dimensionless constants. Equations (\ref{SP1}) and (\ref{SP2}) describe the range of possible gravitational spin tensors that account for the angular momentum exchanged with matter; the aim of this current section is to find a distinguished member of this set, deserving of its physical interpretation.  

We encountered a similar ``superpotential''\footnote{These superpotentials are so called because they are total derivatives. They bear no relation to the homonymous concept from supersymmetric field theory.} freedom when deriving $\tau_{\mu\nu}$ in \cite{Butcher10}, and extinguished it immediately by insisting that the energy-momentum tensor should be free of second derivatives. Unfortunately, this tactic is of no use here: all the terms in $s^\alpha_{\phantom{\alpha}\mu\nu}$ have the same form $ h \partial h$, and so cannot be distinguished from one another by their differential structure. Instead, we must place \emph{algebraic} requirements on the spin tensor, and we shall do so by choosing two conditions that are physically well-motivated, and closely analogous to the algebraic properties of $\tau_{\mu\nu}$. 

\subsection{The Plane-wave Condition}
\emph{Condition 1:} The spin tensor of any (harmonic gauge) gravitational plane-wave (\ref{planewave}), with wave-vector $k^\mu$, must obey
\begin{align}\label{c1}
s^\alpha_{\phantom{\alpha}\mu\nu}\propto k^\alpha.
\end{align}
Clearly, this ensures that spin flows in the direction of propagation of the wave, a physically reasonable request that reciprocates the property $\tau_{\mu\nu}\propto k_\mu k_\nu$ of plane-wave energy-momentum. Substituting (\ref{planewave}) into equations (\ref{SP1}) and (\ref{SP2}), we find that the condition (\ref{c1}) holds for all harmonic gauge plane-waves if and only if
\begin{align}\label{c1=}
A_2=- A_1, \quad A_4=A_1 /4, \quad A_3=A_5=0.
\end{align}
This leaves us with a much smaller range of spin tensors
\begin{align}\label{c1s}
\kappa s^\alpha_{\phantom{\alpha}\mu\nu} = h_{\beta[\nu}\partial^\alpha h_{\mu]}{}^\beta + A_1\partial_\beta\left(\hb^\alpha {}_{[\mu}\hb^\beta {}_{\nu]}\right),
\end{align}
parametrized by $A_1$. 

Of course, the influence of this first condition is not limited to gravitational plane-waves. In fact, the restriction (\ref{c1s}) automatically endows the spin tensor with two highly desirable properties that apply to much more general gravitational fields. Furthermore, one can check that these two properties occur \emph{only if} the spin tensor takes the form (\ref{c1s}); hence the logic can be reversed, with both properties taken together as conditions on $s^\alpha_{\phantom{\alpha}\mu\nu}$, and (\ref{c1}) derived as a consequence.

\emph{Property 1a:} The spin carried by a transverse-traceless gravitational field (\ref{TTcon}) is purely \emph{spatial}:
\begin{align}\label{spatials}
h_{0\alpha}=0,\ h=0,\ \partial_ih_{ij}=0 \quad \Rightarrow \quad s^\alpha_{\phantom{\alpha}0i}=0. 
\end{align} Not only are transverse-traceless fields blessed with positive energy-density and causal energy-flux, now we see they carry only standard spatial spin! This result is akin to the Frenkel condition \cite{Frenkel} that constrains the spin-tensor of a Weyssenhoff fluid \cite{Weyss, Obuk}: $ S^\alpha_{\phantom{\alpha}0i}=0$, in the rest-frame of the fluid.\footnote{The Weyssenhoff fluid is simply a perfect fluid with intrinsic spin. Note that the massive spin-1/2 field (described by the classical Dirac Lagrangian) also obeys the Frenkel condition, if one takes the charge current-density to define the field's 4-velocity \cite{Bogo}.} The only difference here is that the gravitational field, being massless, has no rest-frame; in its place, the \TT-frame defines the space/time split.

The reader should not be under the impression that the non-spatial spins $s^\alpha_{\phantom{\alpha}0i}$ are completely unphysical, however; as a matter of fact, they have a simple physical interpretation. In section \ref{MoE}, we explain that the non-spatial angular momentum current-densities $j_{0i}^{\phantom{0i}\alpha}$ localize gravity's \emph{Moment-of-Energy}, the conserved quantity associated with the symmetry of the background under Lorentz boosts. Accordingly, the intrinsic current-densities $s^\alpha_{\phantom{\alpha}0i}$ signify an ``internal displacement of energy'' of the field. This alters the gravitational Moment-of-Energy just as the ``internal spinning motion'', signified by $s^\alpha_{\phantom{\alpha}ij}$, contributes to the total gravitational angular momentum. Due to (\ref{spatials}) it is now clear that the transverse-traceless field does not carry these internal displacements, and hence, that the location of gravitational energy is determined by $\tau_{\mu\nu}$ alone. The $s^\alpha_{\phantom{\alpha}0i}$ still play an important role in the local exchange of Moment-of-Energy with matter (see \S\ref{MoEexchange}) because \TT-gauge cannot be adopted where $\check{T}_{\mu\nu}\ne0$. 

\emph{Property 1b:} All \emph{static} distributions of matter give rise to  \emph{spinless} gravitational fields:\footnote{In order that the distribution does not collapse under its own gravity, the matter will also have stresses $T_{ij} \sim O(\rho h)\sim O(h^2)$, but these can be neglected in the linear approximation.}
\begin{align}\label{static}
\check{T}_{\mu\nu}=  \rho(\vec{x})  \delta^0_\mu \delta^0_\nu\quad \Rightarrow \quad s^\alpha_{\phantom{\alpha}\mu\nu}=0.
\end{align}
The meaning of this statement is intuitively obvious: matter must be \emph{in motion} if it is to generate gravitational intrinsic spin. It is worth remembering that the linearized gravitational field due to static matter is just the Newtonian potential $\Phi$, so (\ref{static}) is equivalent to the statement that $\Phi$ has no spin. This is in keeping with our observation in \cite{Butcher10} that the gravitational field corresponding to a static Newtonian potential has the energy-momentum tensor of a massless scalar field. 

\subsection{The Traceless Condition}
So far we have placed one algebraic condition on the gravitational spin tensor and removed all but one of the superpotential degrees of freedom. Our second condition will fix $A_1$ and determine $s^\alpha_{\phantom{\alpha}\mu \nu}$ uniquely.

We begin by noting that (as shown in our subsequent paper \cite{Ang2}) the superpotential $\partial_\beta(h^\alpha {}_{[\mu}h^\beta {}_{\nu]})$ associated with $A_1$ plays a distinguished role in the \emph{quadratic} approximation to the Einstein field equations. To avoid a major diversion, let us make the following statement here and postpone its proof until \cite{Ang2}: if the difference between the physical metric and the background metric $\phi^*g_{ab}- \gb_{ab}$ is a \emph{local} function of the gravitational field $h_{ab}$, then the quadratic approximation to the vacuum Einstein field equations can be written as
\begin{align}\label{Gh=t}
\widehat{G}_{\mu\nu}^{\phantom{\mu\nu}\alpha\beta}h_{\alpha\beta} = \kappa \left[  \tau_{\mu\nu} + \partial_\alpha (s_{\mu\nu}{}^{\alpha} + s_{\nu\mu}{}^{\alpha} - s^\alpha_{\phantom{\alpha}\mu\nu})/2\right],
\end{align}
if and only if 
\begin{align}\label{A1}
A_1=-1.
\end{align}
The tensor in square brackets on the right-hand side of (\ref{Gh=t}) is the  \emph{Belinfante energy-momentum tensor} of the gravitational field, combining $\tau_{\mu\nu}$ and $s^\alpha_{\phantom{\alpha}\mu \nu}$ into a single object. According to (\ref{Gh=t}) this tensor acts as a source for the gravitational field, fulfilling the role played by the (Belinfante) energy-momentum tensor of matter $T_{\mu\nu}$ at linear order in the non-vacuum equations (\ref{FEqs}). If we wish to be able to interpret $\tau_{\mu\nu}$ and $s^\alpha_{\phantom{\alpha}\mu \nu}$ as a genuine localisation of gravitational energy-momentum and spin, then not only must they (i) account for the exchange of energy-momentum and angular momentum with matter, but  they must also (ii) generate gravity alongside matter in the Einstein field equations; for this reason we must insist that the remaining superpotential freedom be extinguished by setting $A_1=-1$. Were any other value of $A_1$ to be chosen, then a \emph{non-local} field redefinition $h_{\mu\nu}\to h_{\mu\nu}+ O(h^2)$ would be needed to bring the field equations into the form (\ref{Gh=t}), and the physical metric would no longer be a local function of $h_{\mu\nu}$.

Setting $A_1=-1$ also has an important effect on the spin tensor (\ref{c1s}) independent of its role in the quadratic field equations. Once the gauge has been fixed, and $h_{\mu\nu}$ is transverse-traceless, $A_1=-1$ guarantees that the trace of the spin tensor will vanish: $s^\alpha_{\phantom{\alpha}\alpha \nu}=0$. In fact, because the spin tensor (\ref{c1s}) has this property \emph{only if} $A_1=-1$, it is possible to fix the final piece of superpotential freedom by placing a second algebraic condition on the spin tensor, as follows.

\emph{Condition 2:} The spin tensor of a transverse-traceless gravitational field (\ref{TTcon}) must be traceless:
\begin{align}\label{traceless}
h_{0\alpha}=0,\ h=0,\ \partial_ih_{ij}=0 \quad \Rightarrow \quad s^\alpha_{\phantom{\alpha}\alpha \nu}=0.
\end{align}
In appendix \ref{cube}, we further explore the physical interpretation of this condition. As we argue therein, the trace of the spin tensor is proportional to an infinite pressure gradient carried by the gravitational field. Thus, on the understanding that such pressure gradients are unphysical and must be avoided, this discussion offers an interesting alternative justification for the traceless condition. It is also worth remarking that the above condition strengthens the similarity between gravitation spin and standard examples of material spin: the spin tensors of the Weyssenhoff fluid \cite{Weyss, Obuk}, and the spin-1/2 field \cite{Bogo}, are also traceless.

By design, the spin tensor (\ref{c1s}) is consistent with Condition 2 if and only if $A_1=-1$; as a result, we arrive at our final formula for the gravitational spin tensor:
\begin{align}\label{spintensor}
\kappa s^\alpha_{\phantom{\alpha}\mu\nu} = 2 \hb_{\beta[\nu}\partial^{[\alpha} \hb_{\mu]}{}^{\beta]}.
\end{align}
This is the \emph{unique} local, quadratic, Lorentz-covariant function of $h_{\mu\nu}$ that accounts for the local exchange of angular momentum with matter (\ref{AMexchange}) in harmonic gauge (\ref{harmonic}), satisfies the two physically well-motivated algebraic conditions (\ref{c1}) and (\ref{traceless}), and contains no dimensionful constants other than $\kappa$.\footnote{This derivation has taken place entirely within harmonic gauge, so it goes without saying that $s^\alpha_{\phantom{\alpha}\mu\nu}$ has only been uniquely defined up to the addition of terms proportional to $\partial^\mu\hb_{\mu\nu}$. As long as we remain in harmonic gauge (which we must if we are to interpret $\tau_{\mu\nu}$ and  $s^\alpha_{\phantom{\alpha}\mu\nu}$ physically) then such terms can clearly be ignored. For a discussion of the unique extension of $\tau_{\mu\nu}$ and  $s^\alpha_{\phantom{\alpha}\mu\nu}$ beyond harmonic gauge, see section II of our subsequent paper \cite{Ang2}.} This is an exceptionally compact formula, and one that embodies a remarkably parsimonious description of gravitational spin: for a transverse-traceless field, $s^\alpha_{\phantom{\alpha}\mu \nu}$ is specified by no more than 9 independent components (due to (\ref{spatials}) and (\ref{traceless})) as opposed to the 24 that would be needed in the generic case.

This completes the foundational portion of the article. Following the structure of \cite{Butcher10}, our next task is to apply our newly assembled framework to an investigation of the angular momentum absorbed by an infinitesimal gravitational detector. Section \ref{Inter} will focus on the exchange of standard (i.e.\ spatial) angular momentum  $j_{ij}^{\phantom{ij}\alpha}$, and the microaverage that renders this process gauge-invariant; section \ref{MoE} concerns the interpretation of \emph{non-spatial} angular momentum $j_{i0}^{\phantom{i0}\alpha}$, and the physical consequences of its exchange. A reader whose primary interests are the theoretical underpinnings of $\tau_{\mu\nu}$ and $s^\alpha_{\phantom{\alpha}\mu \nu}$ may wish to skip to \cite{Ang2} at this point: knowledge of sections \ref{Inter}, \ref{MoE}, and \ref{Apps} will not be necessary for the discussion therein.

\section{Angular Momentum Microaverage}\label{Inter}
Having derived the formula (\ref{spintensor}) for gravitational spin, we now possess a complete description of the local energy, momentum, and angular momentum carried by the linear gravitational field. Our first application of this framework will be an analysis of the angular momentum exchanged with an infinitesimal probe. This will allow us to revisit the \emph{microaverage}, the procedure which defined the gauge-invariant energy-momentum transferred onto the probe,\footnote{As our description of gravitational energetics only exists in harmonic gauge, we need only consider gauge transformations which do not break the harmonic condition (\ref{harmonic}). Hence, we use the term \emph{gauge-invariant} to mean \emph{invariant under the gauge freedom that remains after enforcing the harmonic  condition}.} and motivated the (equivalent) program of preparing the incident field in transverse-traceless gauge. Clearly, this gauge-fixing program also provides us with an unambiguous definition of the angular momentum exchanged with the probe. What is not obvious, though, is whether a microaveraging procedure can also achieve this effect, allowing us to define a gauge-invariant exchange of angular momentum that does not rely on gauge-fixing. The aim of this section is to confirm the truth of this idea.

We shall consider a system that is almost identical to the one described in section IV A of \cite{Butcher10}: a point-like detector in the path of a gravitational ``pulse'' plane-wave. The gravitational detector will once again consist of an infinitesimal point-source at $\vec{x}=0$,\footnote{It might appear that we risk a loss of generality in placing the probe at the origin, but this is not the case. To explain, let us consider a uniform translation of the coordinates $x^i \to x^i + a^i$; the probe then lies at $\vec{x}=\vec{a}$, and according to (\ref{t+s}) the only effect on the gravitational angular-momentum current-density is $\Delta j_{ij}^{\phantom{ij}\alpha} = 2a_{[i}\tau_{j]}{}^\alpha$.  Because $a^i$ is constant,  the exchange of angular-momentum associated with this term is simply $\partial_\alpha (\Delta j_{ij}^{\phantom{ij}\alpha}) = 2a_{[i}\partial^\alpha\tau_{j]}{}_\alpha$, and we already know from (\ref{MA=TT}) that $\partial^\alpha \tau_{i\alpha}$ (which quantifies the local exchange of linear momentum) is rendered gauge-invariant by the monopole-free microaverage: $\langle\partial_\alpha (\Delta j_{ij}^{\phantom{ij}\alpha})\rangle^\slashed{M}{\!\!\!\!\!\!}_{\int} =  2a_{[i}\langle\partial^\alpha\tau_{j]}{}_\alpha \rangle^\slashed{M}{\!\!\!\!\!\!}_{\int}$. Clearly, this term accounts for the angular momentum that results from the transfer of linear momentum onto the detector; by assuming that the probe is at $\vec{x}=0$ in what follows, we are simply ignoring the trivial exchange of angular momentum associated with the detector's bulk motion.}
the energy-momentum tensor of which is given by (\ref{pointsource}) as $M,I_{ij},L_{ij}\to 0$.\footnote{We take this limit as the size of the source shrinks to zero. The detector is then a form of generalised ``test-particle'' with negligible self-interaction in comparison to the effect of the external field.} The gravitational field
\begin{align}
h_{\mu\nu} = h_{\mu\nu}^\mathrm{wave} + h_{\mu\nu}^\mathrm{source},
\end{align}
is the sum of the incoming gravitational wave,
\begin{align}\nonumber
h_{\mu\nu}^\mathrm{wave}&= A_{\mu\nu} \delta(k_\alpha x^\alpha - t_0),& A_{\mu\nu}&= \mathrm{const.},\\\label{deltapulse}
k_\mu &=(1,-1,0,0), & k^\mu \bar{A}_{\mu\nu}&=0,
\end{align}
 and the field  $h_{\mu\nu}^\mathrm{source}$ generated by the detector,
\begin{align}
\partial^2 \bar{h}_{\mu\nu}^\mathrm{source}= - 2 \kappa \check{T}_{\mu\nu}.
\end{align}
It is important to recognise that the plane-wave (\ref{deltapulse}) is not quite the same as the one we used when defining the energy-momentum microaverage. There, the gravitational wave had the profile of a Heaviside step function $H$, and this brought about an exchange of energy-momentum $\partial^\mu \tau_{\mu\nu} \sim \partial h\partial^2 h\sim \delta(t-t_0) \delta(\vec{x})$ that was confined to an infinitesimal spacetime region over which we could average. The same is not true of angular momentum, however: a step function wave will give rise to a local exchange $\partial_\alpha j_{\mu\nu}^{\phantom{\mu\nu}\alpha}$ including a term $\partial_\alpha s^\alpha_{\phantom{\alpha}\mu\nu} \sim h \partial^2 h\sim H(t-t_0)\delta(\vec{x})$ that is not localized at $t=t_0$ and is therefore unsuitable for microaveraging. We have no choice but to use a delta-function wave to generate a point-like angular momentum exchange.\footnote{One might try to use a pulse based on derivatives of the delta-function, but the process of splitting a general wave into such pulses is non-local and introduces an arbitrary constant of integration.} This will be the only modification needed to adapt the microaverage for angular momentum.\footnote{The lesson here is that the microaverage is not a process in which we split the incident wave into a particular sort of pulse: as we have seen, the profile of the pulse depends on what exchange we are microaveraging. Rather, it is a process in which we split the wave such that the local exchange (of energy-momentum $\partial_\alpha \tau^\alpha_{\phantom{\alpha}\mu}$, or angular momentum $\partial_\alpha j_{ij}{}^{\alpha}$) takes a particular form: a series of delta-function pulses (and possibly derivatives of delta-functions) each of which can then be averaged over a vanishingly small 4-volume.}

Following the same reasoning that took us to equation (42) of \cite{Butcher10}, we find that the exchange of spatial angular momentum for this system is given by
\begin{align}\nonumber
\partial_\alpha j_{ij}^{\phantom{ij}\alpha} &= 2 x_{[i} \partial^\alpha \tau_{j]\alpha} + h^\beta{}_{ [j}\partial^2 \hb_{i]\beta}/\kappa\\\nonumber
&= -\half k_{[j}x_{i]}\dot{\delta}(k_\alpha x^\alpha - t_0)\left[\ddot{I}_{kl}A_{kl}\delta(\vec{x})\right.\\\nonumber
&\qquad\qquad\quad \ \!\!\!{}- 2\big(\dot{I}_{kl}- L_{kl}\big)A_{k0}\partial_l \delta(\vec{x})\\\nonumber
&\qquad\qquad\quad \ \!\!\!{}+ \big(2M\delta(\vec{x}) +I_{kl}\partial_k\partial_l \delta(\vec{x})\big)A_{00} \Big]\\\nonumber
& \quad {}- \delta(k_\alpha x^\alpha - t_0)\left[ A_{0[i}\big(\dot{I}_{j]k} - L_{j]k}\big)\partial_k \delta(\vec{x}) \right.\\\label{premicro}
&\qquad\qquad\!\!\!\qquad\qquad\qquad\qquad \left.{} -A_{k[i}\ddot{I}_{j]k}\delta(\vec{x})\right]\!.
 \end{align}
As was the case with energy-momentum, the local exchange of angular momentum (\ref{premicro}) is clearly not invariant under the gauge transformations
\begin{align}\label{planegauge}
\delta A_{\mu\nu} &= E_{(\mu}k_{\nu)},& E_\mu&=\mathrm{const.},
\end{align}
which neither break the harmonic condition (\ref{harmonic}) nor alter the form (\ref{deltapulse}) of the wave. This gauge dependence can be dealt with in one of two ways. The simplest approach is to invoke the familiar \TT\ program, insisting that the incident field be transverse-traceless: $A_{\mu\nu}= A^\TT_{\mu\nu}$. The alternative, which we will now examine, is to integrate over the infinitesimal interaction region and render the exchange gauge-invariant \emph{without gauge-fixing}. The two methods give identical results, as we shall soon show.

The microaverage $\langle \ldots \rangle_{t_0}$ is defined, just as it was in \cite{Butcher10}, by
\begin{align}\label{microaverage}\nonumber
\langle f\rangle_{t_0}&\equiv\delta(\vec{x})\delta(t-t_0) \lim_{\epsilon \to 0} \int_{\mc{B}_\epsilon (t_0)} f \ud^4 x,\\ \ \text{where} \quad \mc{B}_\epsilon(t_0)& \equiv \{(t,\vec{x}):  |t - t_0| \le \epsilon,|\vec{x}| \le \epsilon\}.
\end{align}
Applying this definition to (\ref{premicro}) and integrating by parts,\footnote{For each term, integrate by parts to move derivatives from $\delta(\vec{x})$ onto the $x_i\dot{\delta}(k_\alpha x^\alpha - t_0)$ or  $\delta(k_\alpha x^\alpha - t_0)$ part of the term, convert $\partial_i \delta(k_\alpha x^\alpha - t_0) = - \delta_{1i} \dot{\delta}(k_\alpha x^\alpha - t_0) $, and integrate by parts once again to send the time-derivatives to the $M,J_{ij}, I_{ij}$ part of the term, recalling that $\dot{M}=\dot{J}_{ij}=0$. Note that at least one of the spatial derivative must act on the $x_i$ in front of the orbital terms: those terms where $x_i$ is left untouched will vanish because $\delta(\vec{x})$ will set $x_i=0$ when the integral is finally evaluated.} we arrive at 
\begin{align}\nonumber
\langle \partial_\alpha j_{ij}^{\phantom{ij}\alpha}\rangle_{t_0}&=\delta(\vec{x})\delta(t-t_0)\left[k_{[j}\big(\ddot{I}_{i]k}A_{k0} +  \ddot{I}_{i]1} A_{00}\big) \right.\\
&\quad\left.{}+ A_{0[i}\ddot{I}_{j]1} +A_{k[i}\ddot{I}_{j]k}\right].
\end{align}
Although it is far from obvious in its current form, this equation is in fact invariant under the gauge transformations given in equation (\ref{planegauge}). The easiest way to demonstrate this is to examine each component in turn and to use $k^\mu \bar{A}_{\mu\nu}=0$ in the following form:
\begin{align}\nonumber
A_{00}+ A_{11} +2 A_{01} &=0,& A_{22}+A_{33}&=0,\\ \label{unpack}
A_{02} +A_{12}&=0,& A_{03} +A_{13}&=0.
\end{align}
After a great deal of canceling, one finds that
\begin{align}\nonumber
\langle \partial_\alpha j_{23}^{\phantom{23}\alpha}\rangle_{t_0}&=
\delta(\vec{x})\delta(t-t_0)\left(A_+ \ddot{I}_{23}- A_\times
(\ddot{I}_{22}- \ddot{I}_{33})/2\right),\\\nonumber
\langle \partial_\alpha j_{12}^{\phantom{12}\alpha}\rangle_{t_0}&= -
\delta(\vec{x})\delta(t-t_0)\left(A_+ \ddot{I}_{12} + A_\times
\ddot{I}_{13}\right)/2,\\\label{microj}
\langle \partial_\alpha
j_{13}^{\phantom{13}\alpha}\rangle_{t_0}&= -
\delta(\vec{x})\delta(t-t_0)\left(A_\times \ddot{I}_{12} - A_+
\ddot{I}_{13}\right)/2,
\end{align}
all of which depend only on the transverse components of the wave $A_\times= A_{23}$, $A_+=(A_{22}- A_{33})/2$ which are invariant under (\ref{planegauge}). Considering that the microaverage was developed purely for the purposes of \emph{energy-momentum} exchange, it is gratifying to discover that it renders the exchange of angular momentum gauge-invariant as well.

It is possible to write the above relations (\ref{microj}) in a more compact form: 
\begin{align}\label{Amicro}
\langle \partial_\alpha j_{ij}^{\phantom{ij}\alpha}\rangle_{t_0}=
\delta(\vec{x})\delta(t-t_0) A^\TT_{k[i}\ddot{I}_{j]k},
\end{align}
where $A^\TT_{\mu\nu}$ is the transverse-traceless part of $A_{\mu\nu}$, the only non-zero components of which are  $A^\TT_{22}=-A^\TT_{33}=A_{+}$ and $A^\TT_{23}=A^\TT_{32}=A_{\times}$. As previously advertised, this is exactly the same result as would be obtained from applying the \TT\ program to the bare angular momentum exchange (\ref{premicro}):
\begin{align}\nonumber
\partial_\alpha j_{ij}^{\TT \, \alpha}&\equiv \partial_\alpha j_{ij}^{\phantom{ij}\alpha}[h^\text{source}_{\mu\nu}+ A^\TT_{\mu\nu}\delta(k_\alpha x^\alpha -t_0)]\\
\label{ATT}&=\delta(\vec{x})\delta(t-t_0) A^\TT_{k[i}\ddot{I}_{j]k}.
\end{align} 
The only subtlety with this calculation is that one must set $x_i \dot{\delta}(k_\alpha x^\alpha - t_0) \delta(\vec{x})=0$, which is valid as an identity between distributions on test functions that are differentiable with respect to $t$ at $(t_0,\vec{0})$.

The angular momentum microaverage need not be restricted to plane-wave pulses: we can generalise equation (\ref{Amicro}) following the same procedure as the energy-momentum case. First we note that an arbitrary (harmonic-gauge) plane-wave
\begin{align}\label{Bplane}
h^\text{wave}_{\mu\nu}= B_{\mu\nu}(k_\alpha x^\alpha),\qquad k^\mu\bar{B}_{\mu\nu}=0,
\end{align}
can be split into a sum of individual pulses
\begin{align}
h^\text{wave}_{\mu\nu}&= \int^{\infty}_{-\infty} B_{\mu\nu}(t_0)\delta(k_\alpha x^\alpha-t_0)\ud t_0,
\end{align}
and the angular momentum exchange of each pulse microaveraged separately:\footnote{This microaverage carries the subscript ${\int\!\delta}$ to remind us that the wave has been split into $\delta$-function pulses, rather than Heaviside steps.}
\begin{align}\nonumber
&\left\langle\partial_\alpha j_{ij}^{\phantom{ij}\alpha}[h^\text{source}_{\mu\nu}+h^\text{wave}_{\mu\nu}]\right\rangle_{\int\!\delta}\\
\label{intmicro}&\equiv \int^{\infty}_{-\infty}\langle \partial_\alpha j_{ij}^{\phantom{ij}\alpha}[h_{\mu\nu}^\text{source}+ B_{\mu\nu}(t_0) \delta(k_\alpha x^\alpha-t_0)] \rangle_{t_0}\ud t_0.
\end{align}
Second we recall that any incident field $h^\text{in}_{\mu\nu}$ can be expressed as a sum of plane-waves, at least locally. Because (\ref{Amicro}) is linear in the incident field, we can split any incident field into a sum of plane-waves, each of which can be split into a sum of pulses, then perform the microaverage on each element and reassemble the result. The general formula is therefore
\begin{align}\label{hinTT}
\langle \partial_\alpha j_{ij}^{\phantom{ij}\alpha}\rangle_{\int\!
\delta} = \delta(\vec{x}) h^\TT_{k[i}\ddot{I}_{j]k},
\end{align}
where $h^\TT_{\mu\nu}$ is the transverse-traceless part of $h^\text{in}_{\mu\nu}$. 

This concludes our analysis of the spatial angular momentum transferred onto the probe. The non-spatial currents $j_{0i}^{\phantom{0i}\alpha}$ can also be absorbed by the detector; the exchange equation (\ref{AMexchange}) then ensures that the shift in gravity's moment-of-energy is accompanied by a displacement in the detector's centre-of-mass. This is a rather surprising phenomenon, and one that, to our knowledge, has not been discussed in the literature. Under resonant conditions, this effect can cause the detector to ``walk'' in a direction transverse to the gravitational wave.\footnote{This should not be confused with the motion associated with the linear momentum that the probe gains according to (\ref{MA=TT}). There, a resonance between the detector and the incident wave gives rise to a longitudinal acceleration, and the velocity gained in this process remains after the has wave passed.} The next section is devoted to a detailed examination of this phenomenon.

\section{Moment of Energy}\label{MoE} 
Through its unification of space and time, and energy and momentum, special relativity fused together the once disparate notions of angular momentum and centre-of-mass. In this section we review this idea in terms of local currents, and offer an interpretation for the non-spatial intrinsic spin currents  $s^\alpha_{\phantom{\alpha}0i}$ . We also examine the local exchange of moment-of-energy between the gravitational field and an infinitesimal detector. In appendix \ref{FirstP} we confirm that this phenomenon is also predicted by a ``first principles'' description of the system.

\subsection{Definitions and Interpretation}
It goes without saying that the non-spatial components $j_{0i}^{\phantom{0i}\alpha}$ and $J_{0i}^{\phantom{0i}\alpha}$ are needed to form the Lorentz-covariant currents $j_{\mu\nu}^{\phantom{\mu\nu}\alpha}$ and $J_{\mu\nu}^{\phantom{\mu\nu}\alpha}$; thus, at the most basic level, these non-spatial components carry the interpretation of standard ``spatial'' angular-momentum as seen by a moving observer. Beyond this, the non-spatial components carry an additional interpretation that is quite distinct from spatial angular momentum. They are the current densities of a conserved 3-vector quantity: \emph{the Moment-of-Energy}.

To explain, let us first define the moment-of-energy $X_i$, total linear momentum $P_i$, and total mass/energy $M$ for matter:
\begin{align}\nonumber
X_i& \equiv -\int\sqrt{-g} T^0_{\phantom{0}0}y_i\ud^3y, &P_i &\equiv \int\sqrt{-g}  T^0_{\phantom{0}i}\ud^3y, \\
M &\equiv- \int\sqrt{-g} T^0_{\phantom{0}0}\ud^3y,
\end{align}
noting that the centre-of-mass $x^{(0)}_i$ is simply the moment-of-energy normalised by the total mass/energy:
\begin{align}\label{com0}
x_i^{(0)}\equiv X_i/M.
\end{align}
The total non-spatial angular momentum of matter is then
\begin{align}\nonumber
\int \sqrt{-g} J_{0i}^{\phantom{0i}0} \ud^3y& \equiv \int \sqrt{-g}( T^0_{\phantom{0}i}y_0 - T^0_{\phantom{0}0}y_i) \ud^3y \\\label{J1}
&\equiv - tP_i + X_i,
\end{align}
where we have written $y^0=t$.\footnote{Note that we use the same symbol $t$ to represent the value of the time coordinate $y^0$ in physical spacetime and the time coordinate $x^0$ of the background. This has the advantage of allowing us to drop the distinction between the physical quantities $X_i$, $P_i$, $M$, $x_i^{(0)}$, and their background representations $\phi^*(X_i)$, $\phi^*(P_i)$, $\phi^*(M)$, $\phi^*(x^{(0)}_i)$: the first set are functions of $y^0$ only, the second set of $x^0$ only, and the two sets are numerically equal when $x^0=y^0$.}  In the absence of the gravitational field ($h_{\mu\nu}=0$) the angular momentum currents are conserved, 
\begin{align}
\partial_\alpha (\sqrt{-g} J_{\mu\nu}^{\phantom{\mu\nu}\alpha})=\sqrt{-g} \nabla_a J_{\mu\nu}^{\phantom{\mu\nu}a}=0,
\end{align}
and as a result, 
\begin{align}
\partial_t \left( X_i - t P_i \right)=0.
\end{align}
Furthermore, the conservation of energy-momentum ($\partial_\alpha(\sqrt{-g} T^{\alpha}_{\phantom{\alpha} \mu})=\sqrt{-g} \nabla_a J_\mu^{\phantom{\mu}a}=0$) ensures that $\dot{P}_i=0$, and leads to the following global conservation law:
\begin{align}\label{X-P}
\dot{X}_i - P_i =0.
\end{align} 
This equation integrates to $X_i = tP_i + X_i|_{t=0}$, which on substitution into (\ref{J1}) gives
\begin{align}
\int \sqrt{-g}J_{0i}^{\phantom{0i}0} \ud^3y = X_i|_{t=0},
\end{align}
which is constant by definition. In other words, the total non-spatial angular momentum is equal to the \emph{moment-of-energy at} $t=0$, a conserved quantity which we will refer to by the acronym MoE, where the stipulation ``at $t=0$'' should be taken as given.

The same analysis can be performed for the gravitational field in the absence of matter. Working in the background, we define 
\begin{align}\nonumber \label{j0idefs}
\mc{X}^\tau_i &\equiv \int \tau_{00}x_i\ud^3x,& \mc{P}_i& \equiv \int \tau^0_{\phantom{0}i}\ud^3x,\\
\mc{X}^s_i& \equiv \int s^0_{\phantom{0}0i}\ud^3x,& \mc{X}_i&\equiv \mc{X}^\tau_i + \mc{X}^s_i.
\end{align}
Then the total non-spatial gravitational angular momentum is given by
\begin{align} \label{gravX-tP}
\int j_{0i}^{\phantom{0i}0} \ud^3x \equiv \mc{X}_i- t \mc{P}_i, 
\end{align} which, due to $\partial_\alpha j_{0i}^{\phantom{0i}\alpha}=0$ and $\partial_\alpha \tau^{\alpha}_{\phantom{\alpha} i}=0$, is conserved: 
\begin{align}\label{X-Pgrav}
\dot{\mc{X}}_i - \mc{P}_i &=0,\\
\int j_{0i}^{\phantom{0i}0} \ud^3x &= \mc{X}_i|_{t=0}.
\end{align} 
We conclude from this that the $j_{0i}^{\phantom{0i}\alpha}$ are the current-densities of the conserved quantities $\mc{X}_i|_{t=0}$ that constitute the gravitational MoE.

As (\ref{j0idefs}) makes clear, the non-spatial spin densities $s^0_{\phantom{0}0i}$ shift the gravitational MoE by $\mc{X}^s_i$, displacing it from the value $\mc{X}^\tau_i$ that would have been expected from $\tau_{00}$ alone. This suggest that the $s^0_{\phantom{0}0i}$ represent an ``internal displacement of energy'' at a point (analogous to the notion of $s^0_{\phantom{0}ij}$ as ``internal spinning motion'' at a point) so that the field's energy lies locally off-centre. The value of $\tau_{00}(p)$ still represents the density of gravitational energy at the point $p$, but an asymmetry in the distribution of the energy ``within the point'', quantified by $s^0_{\phantom{0}0i}$, shifts the MoE by a small amount.\footnote{This pointwise internal structure (spinning motion and displacements) presumably takes place in the tangent space of the manifold, where the gravitational field is defined.} Because $s^0_{\phantom{0}0i}=0$ for any transverse-traceless gravitational field, these internal displacements rarely arise when describing the energetics of the gravitational field in vacuum. However, as \TT-gauge cannot be adopted where $\check{T}_{\mu\nu}\ne0$, the $s^0_{\phantom{\alpha}0i}$ inevitably play an active role in the exchange of MoE between matter and gravity.

\subsection{Moment of Energy Exchange}\label{MoEexchange}
When matter and gravity interact, neither $j_{0i}^{\phantom{0i}\alpha}$ nor $J_{0i}^{\phantom{0i}\alpha}$ are independently conserved, and MoE is exchanged between them according to (\ref{AMexchange}). Consequently, the conservation laws (\ref{X-P}) and (\ref{X-Pgrav}) are broken,
\begin{align}
\dot{X}_i - P_i&\equiv \Delta \dot{X}_i\ne0,\\
\dot{\mc{X}}_i - \mc{P}_i&\equiv \Delta \dot{\mc{X}}_i\ne0,
\end{align}
but the extent to which they are broken is exactly equal and opposite:\footnote{This global exchange equation follows directly from the local exchange equations: multiply equation (\ref{AMexchange}) by $\phi^*(\sqrt{-g})=\sqrt{-\gb}+O(h)$, discard terms $O(h^3)$, and integrate over the spatial coordinates. This gives $\Delta \dot{\mc{X}}_i + \Delta \dot{X}_i - t (\dot{\mc{P}}_i + \dot{P}_i)=0$, and $\dot{\mc{P}}_i + \dot{P}_i=0$ follows from the  local exchange of linear-momentum (\ref{exchange}) by exactly the same method.}
\begin{align}\label{Xexchange}
\Delta \dot{\mc{X}}_i + \Delta \dot{X}_i = 0.
\end{align}

To understand this process in general, we turn once again to our preferred testing ground: an infinitesimal detector in the path of a gravitational plane-wave. Unlike our analysis of angular momentum for this system (\S\ref{Inter}) we will not employ the microaverage here. The reason for this is simple: the microaverage does not produce a gauge-invariant description of the exchange of MoE. In contrast to angular momentum and energy-momentum, the gauge-invariant modes of the gravitational field do not deliver MoE evenly across the whole detector, they are biased by a dipole term proportional to $\partial_i \delta(\vec{x})$.\footnote{This can be seen in equation (\ref{trans}) below.} The microaverage is therefore unable to capture the exchange properly, as it can only produce quantities proportional to $\delta(\vec{x})$. This is a notable qualitative difference between the exchange of angular momentum and MoE, but in reality it poses no practical difficulty: we can still remove the gauge dependence by insisting that the incident field is transverse-traceless.

With this in mind, we consider the same system as described in section \ref{Inter} with one exception: the incident field is an arbitrary transverse-traceless plane-wave,
\begin{align}\label{Bwavee}
h^\text{wave}_{\mu\nu}= B^\TT_{\mu\nu}(t- x_1),\qquad B^\TT_{0\nu}=B^\TT_{1\nu}=B^{\TT\!}=0,
\end{align}
rather than a pulse. Taking the same steps that were used to derive (42) of \cite{Butcher10}, and deploying the distributional identity $x_i \delta(\vec{x})=0$, we find that the local exchange of non-spatial angular momentum is
\begin{align}\label{long}
\partial_\alpha j_{10}^{\phantom{10}\alpha}&= t \partial^\alpha \tau_{\alpha 1,}\\\label{trans}
i=2,3:\quad \partial_\alpha j_{i0}^{\phantom{i0}\alpha}&= B^\TT_{i k} \left( \dot{I}_{kj} - L_{kj}\right)\partial_j \delta(\vec{x})/2.
\end{align}
As the longitudinal (\ref{long}) and transverse (\ref{trans}) equations represent two very different phenomena, we shall examine them separately.

Equation (\ref{long}) is essentially trivial: it accounts for the extra MoE that arises from the exchange of linear momentum in the $x^1$ direction. To demonstrate this, let us take the time derivative of (\ref{gravX-tP}):
\begin{align}
\dot{\mc{X}}_i - \mc{P}_i-t \dot{\mc{P}}_i = \int \partial_0 j_{0i}^{\phantom{0i}0}\ud^3 x = \int \partial_\alpha j_{0i}^{\phantom{0i}\alpha}\ud^3 x.
\end{align}
Unlike the non-interacting case, we now have
\begin{align}
\dot{\mc{P}}_i = \int \partial_0 \tau^0_{\phantom{0}i} \ud^3 x = \int \partial_\alpha \tau^\alpha_{\phantom{\alpha}i} \ud^3 x,
\end{align}
which is nonzero in general. Consequently,
\begin{align}
\Delta \dot{\mc{X}}_i \equiv \dot{\mc{X}}_i - \mc{P}_i= \int t \partial_\alpha\tau^\alpha_{\phantom{\alpha}i} +\partial_\alpha j_{0i}^{\phantom{0i}\alpha}\ud^3 x .
\end{align}
Thus, the quantity that describes the local exchange of MoE is in fact the sum
\begin{align}
t \partial_\alpha\tau^\alpha_{\phantom{\alpha}i} +\partial_\alpha j_{0i}^{\phantom{0i}\alpha},
\end{align}
as it is this combination which contributes the extra increase in $\mc{X}_i$ beyond what would be expected from simply integrating $\mc{P}_i(t)$ with respect to time. Because the gravitational wave only deposits momentum in the longitudinal direction (see equation (\ref{MA=TT})) this argument has no effect on the interpretation of (\ref{trans}); however, equation (\ref{long}) reveals that
\begin{align}
 t \partial_\alpha\tau^\alpha_{\phantom{\alpha}1} +\partial_\alpha j_{01}^{\phantom{01}\alpha}= t \partial_\alpha\tau^\alpha_{\phantom{\alpha}1}+(-t \partial_\alpha\tau^\alpha_{\phantom{\alpha}1}) =0,
\end{align}
confirming that there is no exchange of MoE in the $x^1$-direction, only the exchange of linear momentum. The center-of-mass of the detector will accelerate in the $x^1$-direction, but this acceleration will be exactly what one would expect from the linear momentum transfer discussed in \cite{Butcher10}.

In comparison, the exchange of transverse MoE (\ref{trans}) is considerably less trivial. The first complication is that $\partial_\alpha j_{i0}^{\phantom{10}\alpha}\propto \partial_j \delta(\vec{x})$, indicating that the transfer of MoE occurs within a dipole-like distribution, taking opposite signs at opposite ends of the detector. In general,  these effects will partially cancel each other, so a more pertinent quantity to calculate (rather than the local exchange) is the \emph{total} MoE exchange over the whole detector:
\begin{align}\nonumber
\Delta \dot{X}_i=-\Delta \dot{\mc{X}}_i& =-\int t \partial_\alpha\tau^\alpha_{\phantom{\alpha}i} +\partial_\alpha j_{0i}^{\phantom{0i}\alpha}\ud^3 x\\\label{totalMoEex}
&=\dot{B}^\TT_{i k}(t) \left( \dot{I}_{k1} - L_{k1}\right)/2,
\end{align}
for $i=2,3$. This equation describes the transverse drift in gravitational MoE, and via (\ref{Xexchange}), the opposite drift in the matter MoE. 

In general, the centre-of-mass of the detector (\ref{com0}) will move according to
\begin{align}\label{com}
\dot{x}_i^{(0)}&=(\Delta\dot{X}_i + P_i)/M - X_i \dot{M}/M^2,
\end{align}
under the influence of the gravitational wave. Focusing our interest on the transverse directions (for which $P_i=0$ for all time) we note that the last term in (\ref{com}) is the product of two small quantities ($X_i$ and $\dot{M}$) and can therefore be neglected in comparison to the first term, which only contains one small quantity ($\Delta \dot{X}$).\footnote{To argue this more rigorously, suppose that the incident wave has amplitude $B$ and frequency $\Omega$, and that the internal motions of the probe have frequency $\omega$ and amplitude $l$. We require $B\ll 1$ in the linear approximation, $\Omega l \ll 1$ to ensure that the probe is much smaller than the gravitational wavelength, and $\omega l < 1$ so that the internal motions are not superluminal. It follows from (\ref{totalMoEex}) that the first term $\Delta \dot{X}/M \sim B l^2 \Omega\omega$, and from (\ref{MA=TT}) that the second term $X \dot{M}/M^2 = (X/M)( \int \partial^\alpha \tau_{\alpha 0}\ud^3x /M) \sim (X/M) B l^2 \Omega \omega^2$. The factor $X/M =\int \Delta\dot{X} \ud t /M$ can be no larger than $(\Delta\dot{X})_{\text{max}} \Delta t /M \sim B l^2 \Omega\omega \Delta t$, where $\Delta t$ is the duration of the interaction. From this we conclude that the second term $X \dot{M}/M^2 \lesssim B^2 l^4 \Omega^2 \omega^3\Delta t$ is negligible in comparison with the first unless the wave and the probe interact for a very long time $\Delta t\sim (B l^2  \Omega \omega^2)^{-1}$. This becomes completely impossible as the length-scale of the probe $l\to0$.} Making these simplifications, and substituting (\ref{totalMoEex}) into (\ref{com}), we finally arrive at a formula for the transverse motion of the detector's centre-of-mass:
\begin{align}\label{com2}
i=2,3:\quad \dot{x}_i^{(0)} =\dot{B}^\TT_{i k} \left( \dot{I}_{k1} - L_{k1}\right)/2M.
\end{align}

It is important to realise that this motion is not simply a ``coordinate effect''. If we were to place a free particle at rest at the origin, then because the plane-wave is \TT, this reference point will remain at $\vec{x}=0$ indefinitely. Equation (\ref{com2}) therefore predicts the displacement of the centre-of-mass relative to this reference point, and the proper distance between the two points will be, to lowest order, equal to the Euclidean distance in the background.

In passing we also note that, when $I_{ij}=0$, the acceleration of the centre-of-mass is exactly that of a spinning test-particle (of mass $M$ and angular momentum $L_{ij}$) as predicted by the linearized Papapetrou-Dixon equations \cite{Pap,Dixon} in transverse-traceless gauge: 
\begin{align}\nonumber
M \ddot{x}^{(0)}_i &= -L_{jk}R_{i0 jk}/2 +O(h^2) \\
&=  -L_{k1} \ddot{B}^\TT_{i k} /2 + O(h^2),
\end{align}
where, because the probe begins at rest, we have taken $\dot{x}^{(0)}_i = O(h)$. Thus (\ref{com2}) generalises this equation to include the effect of the  quadrupole moment $I_{ij}$ of the particle. Because this quantity is time dependent, this allows for the possibility of \emph{resonance} between the probe and the wave, the consequence of which we shall explore in the following example. 

\subsubsection*{Example: Rotating Rod}
\begin{figure}
\centering
\includegraphics[scale=.42]{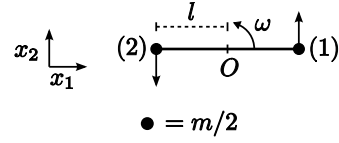}
\caption{A toy model detector: two masses, connected by a light rod, rotate in the $x_3=0$ plane; a gravitational plane-wave, propagating in the $x_1$-direction, disturbs its centre-of-mass in both the longitudinal and transverse directions.}\label{spinner}
\end{figure}

Let us consider the probe depicted in figure \ref{spinner}, a light rod (length $2l$) with bobs of mass $m/2$ at each end, spinning with angular frequency $\omega$ about the $x_3$-axis. A valuable feature of equation (\ref{com2}) is that one only needs the \emph{unperturbed} motion of the detector (as captured by $I_{ij}$ and $L_{ij}$) to calculate the motion of the centre-of-mass to lowest order in $h_{\mu\nu}$; this is not true of a ``first principles'' approach to the problem (see appendix \ref{FirstP}) which complicates that calculation considerably. The unperturbed locations of the two masses are, in the background,
\begin{align}
\vec{x}^{(1)} = l (\cos\omega t, \sin\omega t,0)=-\vec{x}^{(2)},
\end{align} 
and assuming that the speeds are not relativistic (for the sake of simplicity) it is easy to confirm that
\begin{align}\nonumber
\dot{I}_{ij} - L_{ij} =m \omega l^2\left(\begin{matrix}
-\sin(2\omega t )&\cos(2\omega t )-1&0\\\cos(2\omega t) +1&\sin(2\omega t) &0\\0&0&0
\end{matrix} \right) _{ij}.
\end{align}
Inserting this into (\ref{com2}) and setting the total mass/energy $M=m$ under the nonrelativistic assumption, we conclude that centre-of-mass of the spinning rod moves according to
\begin{align}\label{rodcom}
\dot{x}_i^{(0)}&=\frac{\omega l^2}{2} \dot{B}^\TT_{i 2}(t)( \cos(2\omega t) +1),
\end{align}
in the transverse directions $i=2,3$. For a generic gravitational wave, this equation predicts an oscillation in the centre-of-mass that averages to zero over many wavelengths. If the wave is of frequency $2\omega$, however, a resonance occurs in which the detector can steadily ``walk'' in the transverse-direction. A gravitational wave of the form  
\begin{align}
\left(\begin{matrix}
B_{22} \\B_{23} \end{matrix} \right)=  \left(\begin{matrix}
 \beta_+ \\ \beta_\times
\end{matrix} \right) \sin(2\omega (t-x_1)),
\end{align}
gives rise to an average transverse velocity
\begin{align}
\left(\begin{matrix}\label{walking}
\langle \dot{x}_2^{(0)}\rangle \\ \langle \dot{x}_3^{(0)}\rangle
\end{matrix} \right)= \frac{ \omega^2 l^2}{2} \left(\begin{matrix}
\beta_+\\  \beta_\times
\end{matrix} \right).
\end{align}
One of the most surprising aspects of this phenomenon is that the walking motion (\ref{walking}) is not associated with any transverse momentum: $P_2=P_3=0$. The detector moves without being pushed, as it were: due to a careful conspiracy between the probe's internal motion, and the stretching and squeezing of space, the centre of the probe is displaced with each period. 

To understand this on an intuitive level, let us imagine for a moment that the rod joining the masses does not exist, but that at $t=0$ the masses have the same positions and velocities  as before. Because the gravitational wave is invariant under translations in the transverse directions, the transverse momentum (i.e. the transverse components of the momentum covectors) of the two particles will be conserved, and hence the total transverse momentum remains zero. However, the velocity vectors of the masses are related to their conserved momentum covectors by the physical metric $g_{ab}$, which varies in the $x^1$-direction. Thus, because the physical metric differs between the positions of the two masses, while their momenta are equal and opposite, their velocities will not be. In this fashion, a gradient in the gravitational field across the detector can cause a drift in the centre-of-mass of the system. The role of the rod in our detector is simply to apply equal and opposite forces to the masses (again, having no effect on the total transverse momentum) so that once $t= \pi/2\omega$ and the gradient of the gravitational wave across the detector has reversed, the masses are now at the same value of $x^1$, and the drift that has occurred in the first quarter-wavelength will not be undone.

In appendix \ref{FirstP} we substantiate this intuitive picture with a detailed rederivation of equation (\ref{totalMoEex}) from first principles. Not only does this further aid our understanding of the phenomenon, it should assuage any concerns that this unfamiliar effect might simply be an unphysical artifact of our formalism. In fact, the subtlety and complexity of this calculation emphasizes the computational advantage of our approach, not only for MoE, but for angular momentum and energy-momentum also.

\section{Gravitational Plane-Waves}\label{Apps}
As a final exploration of our formula (\ref{spintensor}) for gravitational intrinsic spin, we shall evaluate $s^\alpha_{\phantom{\alpha}\mu\nu}$ for a plane-wave. The motivation for this endeavour is to point out a number of interesting features, and to allow for a comparison with other descriptions of gravitational angular momentum.

A transverse-traceless gravitational plane-wave
\begin{align}\nonumber
h_{\mu\nu}&= h_{\mu\nu}(k_\alpha x^\alpha),& k_\mu&=(1,-1,0,0),\\\label{TT2}
h_{\mu 0}&=h_{\mu 1}=h=0,
\end{align}
has an extremely simple spin tensor:
\begin{align}\nonumber
s^\alpha_{\phantom{\alpha}\mu 0}&=s^\alpha_{\phantom{\alpha}\mu 1}=0,\\\label{splane}
\kappa s^\alpha_{\phantom{\alpha}23}&= k^\alpha (h_\times \dot{h}_+ - h_+ \dot{h}_\times),
\end{align}
where $h_\times=h_{23}$ and $h_+=h_{22}=-h_{33}$ are the transverse components of the wave. As one would expect, $s^\alpha_{\phantom{\alpha}\mu \nu}$ describes transverse spatial spin flowing in the direction of propagation of the wave. Furthermore, the amplitude of $s^\alpha_{\phantom{\alpha}23}$ quantifies the internal spinning motion of the field, as can be seen when we consider a monochromatic wave where the ``plus'' and ``cross'' polarisations differ by a phase $\theta$:
\begin{align}\nonumber
h_+&= A_+ \cos\left(\omega(t-x^1)\right),\\
h_\times&= A_\times \cos\left(\omega(t-x^1)- \theta\right).
\end{align}
In this case, the spin-density is constant over spacetime,
\begin{align}
\kappa s^0_{\phantom{0}23}&= \omega A_\times A_+ \sin\theta,
\end{align}
and is greatest in magnitude when the wave is \emph{circularly polarised}, that is, when $\theta=\pm \pi/2$. Note that a wave with a purely linear polarisation will carry no spin at all.

In \cite{Butcher10} we saw that the energy-momentum tensor of a \TT\ gravitational plane-wave was independent of the timelike vector $u^\mu$  that defines the wave's \TT-frame (\ref{TTcon}). A similar property holds for the spin tensor, but it is complicated by the fact that spin is constrained to be spatial with respect to the \TT-frame, that is, $u^\nu s^\alpha_{\phantom{\alpha}\mu \nu}=0$. As we shall see, the longitudinal and non-spatial spins do transform as the \TT-frame is changed, and in doing so they adapt the spin tensor to obey the spatial constraint for the new $u^\mu$; however, the transverse spatial spin current $s^\alpha_{\phantom{\alpha}23}$ is left invariant. To demonstrate this invariance, we perform a gauge transformation on the field (\ref{TT2}) that maintains its plane-wave form, 
\begin{align}\label{planeGT}
\delta h_{\mu\nu}=\partial_{(\mu}\left(\xi_{\nu)}(k_\alpha x^\alpha)\right)= 2 k_{(\mu} \dot{\xi}_{\nu)},
\end{align}
and note that the spin tensor changes by
\begin{align}\nonumber
\kappa \delta s^\alpha_{\phantom{\alpha}\mu \nu}&= k^\alpha k_{[\mu}\left(   h_{\nu]\beta} \ddot{\xi}^\beta -\dot{h}_{\nu]\beta} \dot{\xi}^\beta\right.\\\label{planeGTs}
&\qquad \ \qquad\left. {}+ k^\beta\left(\dot{\xi}_{\nu]} \ddot{\xi}_{\beta} - \ddot{\xi}_{\nu]}\dot{\xi}_\beta\right)\right),
\end{align} 
confirming that $\delta s^\alpha_{\phantom{\alpha}23}=0$.

Now suppose that the gravitational field (\ref{TT2}) has been transformed to a new \TT-frame, so that in some  other Lorentz coordinate system $\{x^{\mu^\prime}\}$ we have $h_{\mu^\prime 0^\prime}=h_{\mu^\prime 1^\prime}=0$. Then by the same calculation that led us to (\ref{splane}) the transformed spin tensor  $s^{\prime a}_{\phantom{\prime a}bc}$ will obey  $s^{\prime \alpha^\prime}{\!}_{\mu^\prime 0^\prime}=s^{\prime \alpha^\prime}{\!}_{\mu^\prime 1^\prime}=0$ exactly as the original tensor did in the original coordinate system. The only non-zero component of the transformed tensor (in the primed basis) will be  $s^{\prime \alpha^\prime}{\!}_{2^\prime 3^\prime}$, and this quantity will also be gauge-invariant by the same argument we used for $s^\alpha_{\phantom{\alpha}23}$. These two gauge-invariant currents are related by the constant factor $2\Lambda^{[2^\prime}{\!}_{2}\Lambda^{3^\prime]}{\!}_{3}$, where $\Lambda^{\mu^\prime}{\!}_{\nu}$ is the Lorentz transformation between the two coordinate bases:
\begin{align}\nonumber
s^{\alpha}_{\phantom{\alpha}23}= s^{\prime \alpha}_{\phantom{\prime \alpha}23}&= \Lambda^{\mu^\prime}{\!}_{2}\Lambda^{\nu^\prime}{\!}_{3}s^{\prime \alpha}_{\phantom{\prime\alpha}\mu^\prime \nu^\prime}\\&= \left(2\Lambda^{[2^\prime}{\!}_{2}\Lambda^{3^\prime]}{\!}_{3}\right)s^{\prime \alpha}_{\phantom{\prime\alpha}2^\prime 3^\prime}.
\end{align}
This constant of proportionality ensures that $s^{a}_{\phantom{a}bc}$ and $s^{\prime a}_{\phantom{\prime a}bc}$ describe exactly the same  spatial transverse spin current \emph{in either basis}: $s^\alpha_{\phantom{\alpha}23}=s^{\prime\alpha}_{\phantom{\prime\alpha}23}$ and  $s^{ \alpha^\prime}{\!}_{2^\prime 3^\prime}=s^{\prime\alpha^\prime}{\!}_{2^\prime 3^\prime}$. Thus, the only effect of a change in \TT-frame is to re-express the same physical information (the transverse spin current of the wave) in terms of spin that is spatial with respect to a new rest-frame. In the absence of some material body (a detector or a source, for example) the massless gravitational plane-wave cannot define a preferred rest-frame, and so the spatial nature of its intrinsic spin will always have this ambiguity.

As a consequence of this, while a plane-wave region can ``sew together'' two different \TT-frames to form a seamless picture of the propagation of gravitational \emph{energy-momentum} (as described in section III D of \cite{Butcher10}) the same cannot be done for angular momentum: there will always be a discontinuity where the spatial spin of one frame is converted into the spatial spin of the other. Even so, one can construct a gravitational spin \emph{pseudovector}
\begin{align}\label{pseudo}
s^\alpha \equiv \epsilon^{\alpha \lambda\mu\nu} s_{\lambda\mu\nu}/2,
\end{align}
which is truly independent of \TT-frame, and will therefore give a continuous description of gravitational spin within the sewing region. The invariance of $s^\alpha$ follows directly from the totally antisymmetric part of (\ref{planeGTs}): $\delta s_{[\alpha\mu\nu]}=0$. The  physical interpretation of this pseudovector is not immediately clear, but suffice it to say that for a plane-wave, $s^\alpha$ captures only the spin that is linearly independent of the wave-vector $k^\mu$.\footnote{We also note that $s^\alpha$ bears a resemblance to the Pauli-Lubanski pseudovector $S^\alpha \equiv  \epsilon^{\alpha \lambda\mu\nu} P_\lambda L_{\mu\nu}/2$, which characterizes the total spin of a particle or matter field, and reduces in the particle's rest-frame to (mass times) the familiar axial angular-momentum vector of non-relativistic mechanics \cite{Ryder}. } For the plane-wave (\ref{TT2}) we have been studying, the spin pseudovector is
\begin{align}
\kappa s^\alpha &= k^\alpha (h_\times \dot{h}_+ - h_+ \dot{h}_\times),
\end{align} 
capturing all the physically pertinent information of (\ref{splane}) in a completely frame-independent fashion. 

Finally, we should highlight the major difference that exists between the gravitational spin currents in (\ref{splane}) and the corresponding quantities given by the traditional approaches, including the Landau-Lifshitz  tensor \cite{LL} and the integrand of the ADM energy-momentum \cite{ADMII}. In these descriptions, the local energy-momentum and spin of the gravitational field are packaged together in a single object, a \emph{Belinfante} energy-momentum tensor $t_{\mu\nu}\sim \partial h \partial h + h\partial^2 h +O(h^3)$.\footnote{A Belinfante energy-momentum tensor can be constructed from any energy-momentum tensor and spin tensor, including our own: $t_{\mu\nu}[\tau,s] \equiv \tau_{\mu\nu} + \partial_\alpha (s_{\mu\nu}^{\phantom{\mu\nu}\alpha}+s_{\nu\mu}^{\phantom{\mu\nu}\alpha}-s^\alpha_{\phantom{\alpha}\mu\nu})/2$. We perform this calculation in \cite{Ang2} and compare the result with the Landau-Lifshitz and ADM Belinfante tensors discussed here.}  The local angular-momentum currents are then $x_{[\mu}t_{\nu]}{}^{\alpha}$ alone, with no extra ``intrinsic'' component. According to this viewpoint, there is no transverse angular momentum within a harmonic-gauge plane-wave: $x_{[2}t_{3]}{}^{\alpha}=0$.\footnote{This follows from simple index combinatorics. Within the plane- wave $t_{\mu\nu}\sim k k \dot{h} \dot{h} + k k h \ddot{h} + O(h^3)$, and because $k_{\mu}k^\mu=0$ and $k^\mu \bar{h}_{\mu\nu}=0$, both the free indices must occur on the wave-vectors, i.e.\ $t_{\mu\nu}\propto k_\mu k_\nu$. This continues to be true at higher order, where the terms in  $t_{\mu\nu}$ are of the form $kk \dot{h} \dot{h}  h^{n-2} + kk\ddot{h}h^{n-1}  $. Consequently, the transverse angular momentum vanishes exactly: $x_{[2}t_{3]}{}^{\alpha}\propto x_{[2}k_{3]}k^{\alpha}=0$.} This differs dramatically from our description  (\ref{splane})  and stands opposed to the intuitive notion of intrinsic spin as quantifying the internal spinning motion of the field. Without separating gravitational energy-momentum and spin into two separate tensors, $\tau_{\mu\nu}$ and $s^\alpha_{\phantom{\alpha}\mu\nu}$, the intrinsic spin carried by a (harmonic-gauge) plane-wave can never be manifestly present within the wave.

To be clear: the Belinfante-style descriptions still correctly quantify the \emph{total} angular momentum of the wave, but they assign this angular momentum to the wave's boundary, not its interior.\footnote{To avoid a discussion of the boundary at infinity, suppose the plane-wave is in fact restricted to a spatially compact region; in this case, one will find that $x_{[2}t_{3]}{}^{\alpha}\ne0$ at the boundary of the region, and the spatial integral of $x_{[2}t_{3]}{}^{0}$ will amount to the same total angular momentum described by $s^0_{\phantom{0}23}$. In fact, it is generally true that (under suitable boundary conditions) $t_{\mu\nu}[\tau,s]$ gives the same global measure of energy-momentum and angular momentum as $\tau_{\mu\nu}$ and $s^\alpha_{\phantom{\alpha}\mu \nu}$; see \cite{Ang2} for details.} Considering that the angular momentum currents along this boundary are given by $x_{[2}t_{3]}{}^{\alpha}$ as always, and are thus explicitly dependent on $x^\mu$, even these currents cannot be thought of as a local and intrinsic property of the field. This perverse picture, in which all the spin of a gravitational wave resides on the edge of the wave, and this supposedly intrinsic quantity depends on the coordinate distance from the origin, only emphasizes what was already well-known: the Landau-Lifshitz tensor and the integrand of the ADM energy-momentum should not be taken seriously as \emph{local} descriptions of gravitational energy-momentum or spin. While they certainly define meaningful \emph{global} quantities \cite{ADMIVc}, the gauge-freedom of these Belinfante tensors cannot be fixed in a natural manner, and they commonly display negative energy-density and spacelike energy-flux. 

\section{Conclusion}
Together, the energy-momentum tensor $\tau_{\mu\nu}$ and the spin tensor $s^\alpha_{\phantom{\alpha}\mu\nu}$ completely characterize the energy, momentum, and angular momentum carried locally by the linearized gravitational field:
\begin{align}\label{tau2}
\kappa \bar{\tau}_{\mu\nu}&=\tfrac{1}{4}\partial_\mu  h_{\alpha\beta}\partial_\nu \hb^{\alpha\beta},\\\label{spin2}
\kappa s^\alpha_{\phantom{\alpha}\mu\nu} &= 2 \hb_{\beta[\nu}\partial^{[\alpha} \hb_{\mu]}{}^{\beta]}.
\end{align}
The gauge freedom of this description is highly constrained by the harmonic gauge condition,
\begin{align}\label{harm2}
\partial^\mu \hb_{\mu\nu}=0,
\end{align}
which arose as a consequence of the derivation of $\tau_{\mu\nu}$; the last remnant of this freedom is then eliminated by insisting that the incident gravitational field be transverse-traceless, a program motivated in part by appealing to the gauge-invariant exchange of energy-momentum between gravity and an infinitesimal probe, and also distinguished by the numerous desirable properties that the tensors display in transverse-traceless gauge: positive energy-density, causal energy-flux, and spatial spin. 

We developed this framework around a simple principle: wherever the energy, momentum, or angular momentum of matter is changed under the influence of gravity, there must be an equal and opposite change in the energy, momentum, or angular momentum of the gravitational field. This idea, and the requirement that $\tau_{\mu\nu}$ be symmetric and free of second derivatives, was enough to determine the energy-momentum tensor (\ref{tau2}) and the field condition (\ref{harm2}). To determine the spin tensor uniquely, we demanded that it satisfy two physically-motivated conditions: first, the spin of a gravitational plane-wave must flow in the direction of propagation of the wave (\ref{c1}); second, a transverse-traceless field must possess a traceless spin tensor (\ref{traceless}). The latter condition ensures that \emph{local} field redefinitions suffice to cast $\tau_{\mu\nu}$ and $s^\alpha_{\phantom{\alpha}\mu\nu}$ as sources of gravity in a quadratic approximation to general relativity (\ref{Gh=t}) and simultaneously rids the gravitational field of infinite pressure gradients (see appendix \ref{cube}). The resulting spin tensor (\ref{harm2}) displays a number of notable properties that were not required of it: the Newtonian potential has vanishing spin-tensor (\ref{static}) and any transverse-traceless field carries purely spatial spin (\ref{spatials}).

The microaverage, which defines the gauge-invariant exchange of energy-momentum between gravity and an infinitesimal probe, also renders the exchange of spatial angular momentum gauge-invariant (\ref{hinTT}) without the need for gauge-fixing. In the same system, the exchange of non-spatial angular momentum can displace the center-of-mass of the detector, beyond that which would be expected due to the exchange of linear momentum alone (\ref{com2}). Indeed, if the internal motions of the probe resonate with the incident wave, the detector may ``walk'' in a transverse direction, and acquire a net displacement over many wavelengths. We have explored this phenomenon for the specific example of a rotating rod (\ref{rodcom}) and rederived our predictions from first principles (see appendix \ref{FirstP}). 

Unlike $\tau_{\mu\nu}$, the spin tensor of a gravitational plane-wave is not completely independent of the \TT-frame in which the wave is prepared. While the current-density of transverse spatial spin (in any frame) is invariant, the full tensor adapts so as to remain spatial with respect to whichever \TT-frame is used. Thus, if a plane-wave region is used to sew together two \TT-frames and produce a seamless picture of energy-momentum propagation, there will inevitably be a discontinuity in $s^\alpha_{\phantom{\alpha}\mu\nu}$ where the spin is projected from one spatial hypersurface to another; however, a spin psuedovector can be defined (\ref{pseudo}) that is conserved across this interface. 

The spin carried by a plane-wave (\ref{splane}) is also an excellent example with which to compare our framework to the familiar ``Belinfante'' energy-momentum tensors of Landau and Lifshitz, and Arnowitt, Deser and Misner. Whereas $s^\alpha_{\phantom{\alpha}\mu\nu}$  describes spin that is manifestly present within the wave, the density of which depends on the rotational motion of $h_{\mu\nu}$ at each point, the Belinfante tensors assign all angular-momentum to the boundary of the wave, and its density there is not simply a function of $h_{\mu\nu}$ (as a truly local intrinsic property of the field would be) but is also dependent on the distance of the point from the origin. 

Returning to our previous paper \cite{Butcher10}, it becomes clear that many of the remarkable properties of our gravitational energy-momentum tensor (including its positive energy-density and causal energy-flux) owe their existence to the careful separation of gravitational energy-momentum and gravitational spin. Now that we have made this separation explicit, and derived a formula for $s^\alpha_{\phantom{\alpha}\mu\nu}$, we have all the ingredients necessary to understand the broader theoretical picture in which our description resides. This is the task of our next paper \cite{Ang2}, the results of which, in many respects, are the main reward for our work here.

\begin{acknowledgments}
L.\,M.\,B.\ is supported by STFC, St.\,John's College, Cambridge, and Jesus College, Cambridge. We thank Stanley Deser and Richard Arnowitt, without whose helpful comments we may never have begun this investigation into gravitational spin.

\end{acknowledgments}

\appendix

\section{Moment of Energy Exchange from First Principles}\label{FirstP}
In order to rederive equation (\ref{totalMoEex}) from first principles, we shall consider a detector, centred at the origin, composed of a set of $N$ test-particles connected by some form of ``light'' mechanical apparatus.\footnote{The adjective ``light'' is used to indicate that the total energy-momentum of the apparatus is negligible compared to that of the particles.} The $n^{\text{th}}$ particle has mass $m_n$, proper time $\tau_{n}$, and follows a worldline $y_{n}^\mu(\tau_{n})$ in the physical spacetime; its 4-velocity $u_{n}^\mu\equiv \ud y_{n}^\mu/\ud \tau_{n}$ has unit norm: $u_n^{a}u_n^{b}g_{ab}=-1$. In this approach, the detector is not truly infinitesimal, but we stipulate that the length-scale of the detector $l\sim y^i_{n}$ be sufficiently small that we may ignore terms $O(l^3)$ in our calculation, leaving us with a quadrupole approximation of the probe. As usual, a weak gravitational plane-wave is incident upon the detector, represented in transverse-traceless gauge in the background: $h_{\mu\nu}=h^\mathrm{wave}_{\mu\nu}$ as given in (\ref{Bwavee}). As $\partial_2 h_{\mu\nu}=\partial_3 h_{\mu\nu}=0$, the physical spacetime is isometric under translations in the $y^2$ and $y^3$ directions of the $\{y^\mu\}$ coordinate system; for this section, we will reserve the index $k=2,3$ for these transverse directions. 

In the physical spacetime, the energy-momentum tensor of the particles is 
\begin{align}
(T_\mathrm{particles})^a_{\phantom{a}b} &=\sum^N_{n=1} \frac{1}{\sqrt{-g}}\int\ud\tau_{n} \delta(y^\mu -y_n^\mu(\tau_{n}) )u_{n}^a p_{nb},
\end{align}
where $p_{n a}\equiv m_n g_{ab} u_{n}^b$ is the 4-momentum of the $n^{\text{th}}$ particle. For the purposes of defining the moment-of-energy of the detector, we assume that the energy-momentum of the light apparatus is negligible:
\begin{align}\nonumber
X_i& = -\int\sqrt{-g} (T_\mathrm{particles})^0_{\phantom{0}0}y_i\ud^3y\\
&= -\sum^N_{n=1}p_{n0} y_{ni}.
\end{align}
In terms of background quantities, this is
\begin{align}\nonumber
X_i& = -\sum^N_{n=1}m_n (\eta_{0\alpha} + h_{n0\alpha})x_{n}^{\prime \alpha} x_{ni}\\\label{X=xprime}
& =\sum^N_{n=1}m_n x_n^{\prime 0} x_{ni},
\end{align}
where $x_n^\mu(\tau_n)$ are the coordinates of the particles in the background spacetime, primes indicate differentiation with respect to $\tau_{n}$, and $h_{n\mu\nu}\equiv h_{\mu\nu}(x_n^\alpha(\tau_n))$ is the gravitational field evaluated at the $n^{\text{th}}$ particle. The rate of change of the moment-of-energy  is therefore
\begin{align}\label{Xdot1}
\dot{X}_i &= \sum^N_{n=1}m_n\left(\frac{x_n^{\prime\prime0} x_{ni}}{ x_n^{\prime0}} + x^\prime_{ni}  \right).
\end{align}
The normalisation of the 4-velocity,
\begin{align}
-1 = -(x_n^{\prime0})^2 + (h_{nij} + \delta_{ij}) x_{n}^{\prime i} x_{n}^{\prime j},
\end{align}
ensures that $x_n^{\prime0}\sim O(1)$ and $x_n^{\prime\prime0}\sim O(l^2)$, so the first term (\ref{Xdot1}) is $O(l^3)$ and can therefore be neglected. Consequently,
\begin{align}\label{Xddot}
\ddot{X}_i &= \sum^N_{n=1} \frac{m_n x^{\prime\prime}_{ni}}{ x_n^{\prime0}} + O(l^3).
\end{align}

The accelerations $x^{\prime\prime}_{ni}$ in (\ref{Xddot}) are caused by both the gravitational field and the mechanical forces exerted on the particles by the apparatus. Our aim is to infer $\ddot{X}_i$ without assuming any detailed model of the latter. This might seem an impossible task, as it appears that we will need to know the motions of the particles (or the forces from the apparatus) to \emph{first order} in $h_{\mu\nu}$ if we wish to calculate the first order contribution to $\ddot{X}_k$. Fortunately, because the apparatus is light, and the transverse momentum is conserved, only the \emph{unperturbed} motions of the particles will be required. To see this, we start by calculating the linear momentum of the probe:
\begin{align}\nonumber
P_i& = \int\sqrt{-g} (T_\mathrm{particles})^0_{\phantom{0}i}\ud^3y\\
&= \sum^N_{n=1}p_{ni},
\end{align}
where once again we assume that the momentum of the apparatus can be neglected. Because the physical spacetime is isometric under translations in the $y^2$ and $y^3$ directions, the transverse momentum $P_k$ will be conserved:\footnote{This follows from the standard argument: $ \partial_k g_{\alpha\beta}=0$ guarantees that $ (\partial_k)^a$ is a Killing vector, $\nabla^{(a} (\partial_k)^{b)}=0$, and thus $0=\sqrt{-g}\nabla_a(T^a_{\phantom{a} b} (\partial_k)^b)= \partial_\alpha (\sqrt{-g}T^\alpha_{\phantom{\alpha}k})$, the spatial integral of which is $\dot{P}_k=0$.}
\begin{align}\nonumber
0= \dot{P}_k &= \partial_t \left( \sum^N_{n=1}m_n(\delta_{ki} + h_{nki})x^{\prime i}_n \right) \\\label{Fbalance}
&= \sum^N_{n=1}\frac{m_n}{x_n^{\prime0}}\left(x^{\prime\prime}_{nk} +\partial_{\tau_n}( h_{nki}x^{\prime i}_n) \right),
\end{align}
which is equivalent to the statement that the mechanical forces on the particles (due to the apparatus) balance one another.\footnote{Although the total momentum of the apparatus is assumed to be negligible, we have not made any assumptions about the local flux of momentum between the apparatus and the particles, and so the individual mechanical forces on each particle cannot be neglected. The constraint (\ref{Fbalance}) arises because the apparatus has much less mass than the particles, and so any momentum it were to gain would send it off with a very large velocity that would be impossible to maintain while in contact with the particles; in order to stay connected to the particles, the momentum of the apparatus must remain very small, and the forces acting on the apparatus must (approximately) sum to zero.} Substituting this constraint into equation (\ref{Xddot}) we find that
\begin{align}\nonumber
\ddot{X}_k &= -\sum^N_{n=1} \frac{m_n  }{ x_n^{\prime0}} \partial_{\tau_n}( h_{nki}x^{\prime i}_n) +O(l^3)\\
&= \partial_t \left( - \sum^N_{n=1} m_n   h_{nki}x^{\prime i}_n\right) +O(l^3),
\end{align}
which is easy to integrate:\footnote{The constant of integration is set to zero by the initial conditions: the probe is at rest ($\dot{X}_i=0$) before the wave arrives ($h_{\mu\nu}=0$).}
\begin{align}\label{Xdot2}
\dot{X}_k &= - \sum^N_{n=1} m_n   h_{nki}x^{\prime i}_n + O(l^3).
\end{align}
This is the equation we sought: every instance of $x^{\prime i}_n$ is multiplied by $ h_{\mu\nu}$, so  only the unperturbed motions are needed to determine $\dot{X}_k$ to linear order in the gravitational field. 

The last  step is to relate the $h_{nki}$ to the gravitational field at the origin:
\begin{align}\nonumber
h_{n\mu\nu}&= h_{\mu\nu}(t,\vec{0}) + x^i_n\partial_i h_{\mu\nu}(t,\vec{0}) + O(l^2)\\
&= B_{\mu\nu}^\TT(t) - x^1_n \dot{B}_{\mu\nu}^\TT(t) + O(l^2);
\end{align}
as a result, equation (\ref{Xdot2}) becomes
\begin{align}\nonumber
\dot{X}_k &= -  B_{ki}^\TT\left( \sum^N_{n=1} m_n x^{\prime i}_n\right) + \dot{B}_{ki}^\TT\left( \sum^N_{n=1} m_n  x^1_n x^{\prime i}_n\right) + O(l^3)
\\\nonumber
&= -  B_{ki}^\TT\left( \dot{X}_k \right) + \dot{B}_{ki}^\TT\left( \sum^N_{n=1} m_n  x^1_n x^{\prime i}_n\right) + O(l^3)
\\\label{X=Bx}
&=  \dot{B}_{ki}^\TT\left( \sum^N_{n=1} m_n  x^1_n x^{\prime i}_n\right) + O(h^2) +O(l^3).
\end{align}
This simplifies even further when we notice that
\begin{align}\nonumber
\dot{I}_{ij} - L_{ij}& = \partial_t \left(-\int\sqrt{-g} (T_\mathrm{particles})^0_{\phantom{0}0}y_iy_j\ud^3y \right)\\\nonumber&\quad{} -  2 \int\sqrt{-g} (T_\mathrm{particles})^0{}_{[j}y_{i]}\ud^3y
\\\nonumber
&=\partial_t \left(\sum^N_{n=1} m_n x^{\prime 0}_n x_{n}^i x_{n}^j \right) - 2 \sum^N_{n=1} m_n x^{\prime [j}_{n} x_n^{i]}\\\nonumber&\quad{} + O(h)
\\\nonumber
&=2 \sum^N_{n=1} m_n x_{n}^{\prime (i} x_{n}^{j)}  - 2 \sum^N_{n=1}m_n x^{\prime [j}_{n} x_n^{i]}\\\nonumber&\quad{}  + O(h)+ O(l^4)
\\
&= 2 \sum^N_{n=1} m_n  x^j_n x^{\prime i}_n  + O(h)+ O(l^4),
\end{align}
which gives us our final result:
\begin{align}\label{AResult}
\dot{X}_k &= \dot{B}_{ki}^\TT\left( \dot{I}_{i1} - L_{i1}\right)/2 + O(h^2) +O(l^3),
\end{align}
exactly as predicted by equation (\ref{totalMoEex}). 

It should be clear that our formalism provides a much more direct route to this result: one needs only to produce (\ref{trans}) and integrate, a straight-forward operation that lacks the ``insightful'' steps of the first principles calculation, such as invoking conservation of transverse momentum (\ref{Fbalance}) to remove  degrees of freedom from (\ref{Xddot}). However, the moral of this appendix is not simply that our method is more computationally efficient; equally important is the \emph{intuitive} power that our framework confers. Working from first principles, it is hard to imagine that one would have thought to derive (\ref{AResult}) in the first place, as there is no obvious reason to expect that a gravitational wave would produce a transverse motion in the detector's centre-of-mass. In comparison, our unified picture of local gravitational energetics brought this phenomenon to mind as readily as the exchange of energy, momentum, or angular momentum.

\section{Physical Interpretation of the Traceless Condition}\label{cube}
Here we offer an alternative derivation for the traceless condition (\ref{traceless}) which helps to illuminate its physical interpretation. Let us begin by examining the spin tensor of the transverse-traceless gravitational field in detail. The purpose of this analysis is to isolate an algebraic property of $s^\alpha_{\phantom{\alpha}\mu \nu}$ that signifies unphysical behaviour, and then design a condition so that this possibility cannot arise.\footnote{\label{ft} Note that we restrict our attention to the spin of the \emph{transverse-traceless} gravitational field. The spin tensor can only be expected to have a sensible physical interpretation under the same conditions that $\tau_{\mu\nu}$ describes positive energy-density and causal energy-flux, i.e.\ for all \TT-fields, arbitrary (harmonic-gauge) plane-waves, and static fields. It will be trivial to extend Condition 2 to include the last two cases, and since their inclusion does not constrain $A_1$, it is simpler to ignore them in what follows.}

Because the spin of a transverse-traceless field is spatial (Property 1a) we can write
\begin{align}
s^\alpha_{\phantom{\alpha}ij}\equiv s^\alpha_{\phantom{\alpha}k}\epsilon_{kij},
\end{align} 
where $s^\alpha_{\phantom{\alpha}i}$ is the \emph{axial} spin tensor, the current-density of intrinsic spin about the $x^i$-axis. Each component $s_{ij}$ represents the flux of $x^j$-axis spin in the $x^i$-direction; in other words, $s_{ij} \Sigma$ is the torque (along the $x^j$-axis) that acts on a small surface ($x^i=\text{const.}$) of area $\Sigma$.

\begin{figure}
\centering
\includegraphics[scale=.37]{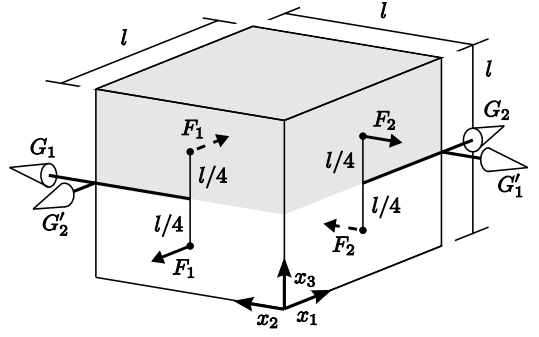}
\caption{The torques on an infinitesimal cube of vacuum due to the flux of gravitational intrinsic spin.}
\label{Cube}
\end{figure}

Let us consider the $l\to 0$ limit of an $l\times l \times l$ cube of vacuum  ($\check{T}_{\mu\nu}=0$, $h_{\mu\nu}\ne0$) as depicted in figure \ref{Cube}. The torque along the $x^2$-axis, acting on the $x^1=0$ face, is $G_1=s_{12}l^2$. There will also be contributions from the $x_{[i}\tau_{j]k}$ part of the angular-momentum current density, but these terms will be of order $l^3$ and so can be safely neglected. It is convenient to think of $G_1$ as being generated by two equal and opposite forces $F_1=2 s_{12} l$ acting on the points $(0, l/2, l/4)$ and $(0, l/2, 3l/4)$ as shown in the diagram. On the opposite face ($x^1=l$) there will be a torque along the $x^2$-axis $G^\prime_1 = -(s_{12} + l \partial_1 s_{12})l^2$, the minus sign arising as a result of the opposite direction of the outward normal, and the second term being negligible as long as $s_{ij}$ is smooth in the cube. Again, this torque can be thought of as being generated by equal and opposite forces $F_1$ acting at the points $(l, l/2, l/4)$ and $(l, l/2, 3l/4)$. Following the same approach, we render the $x^1$-axis torques on the $x^2=0$ and $x^2=l$ faces as forces $F_2= 2 s_{21} l$ acting on the appropriate points on the cube.

We now split the cube along the plane $x^3 = l/2$, and consider the two  ``half-cubes'' separately. The isotropic pressure acting on each half-cube can be evaluated using the formula
\begin{align}\label{pres}
P= \frac{-1}{6} \sum_r \frac{\vec{f}_r \cdot \vec{n}_r }{A_r},
\end{align}
where the index $r$ enumerates the six faces of the half-cube (each with area $A_r$ and outward unit normal $\vec{n}_r$) and $\vec{f}_r$ is the force acting on the $r^\text{th}$ face.\footnote{To confirm the validity of this formula, describe the forces in terms of a stress tensor $\sigma_{ij}$ by writing $f_{ri}= -\sigma_{ij}n_{rj}A_r$. The sum in (\ref{pres}) then becomes $-\sigma_{ij} \sum_r n_{ri} n_{rj} = -\sigma_{ij}(2 \delta_{ij})=-6P$.} For the upper half-cube ($x^3\ge l/2$) both $F_1$ forces are directed inwards, while the two $F_2$ forces are outwardly directed; thus (\ref{pres}) gives
\begin{align}\label{pressure}
P_\text{upper}&=\frac{-1}{6}\frac{ 2F_2-2F_1}{l^2/2}= \frac{4(s_{12}-s_{21})}{3l},
\end{align}
where we have once again ignored the negligible forces, such as $\tau_{33} l^2$ on the $x^3=l$ and $x^3=l/2$ faces. The calculation for the lower half-cube ($x^3\le l/2$) is identical except that the forces $F_1$ point outward and $F_2$ point inward; as a result, $P_\text{lower}=-P_\text{upper}$. Therefore, within the cube we find a pressure gradient
\begin{align}\label{trace0}
\frac{\partial P}{\partial x^3} \approx \frac{P_\text{upper} -P_\text{lower}}{l/2}=  \frac{16(s_{12} -s_{21})}{3 l^2},
\end{align}
which grows without bound as the limit $l\to0$ is taken! The only way to avoid these infinite pressure gradients is to insist that $s_{[ij]}=0$, or equivalently
\begin{align}\label{traces=0}
s^\alpha_{\phantom{\alpha}\alpha\nu}=0.
\end{align}
Hence the traceless condition (\ref{traceless}) can be derived by requiring that the (gauge-fixed) gravitational field be free of infinite pressure gradients.

\bibliography{Spin}
\end{document}